# MULTISCALE MODELING OF PLASTIC DEFORMATION OF MOLYBDENUM AND TUNGSTEN: II. YIELD CRITERION FOR SINGLE CRYSTALS BASED ON ATOMISTIC STUDIES OF GLIDE OF 1/2⟨111⟩ SCREW DISLOCATIONS

R. Gröger[1,2*], V. Racherla[3,4], J. L. Bassani[3] and V. Vitek[1]

[1] University of Pennsylvania, Department of Materials Science and Engineering, Philadelphia, PA 19104, USA
[2] Los Alamos National Laboratory, Theoretical Division, Los Alamos, NM 87545, USA
[3] University of Pennsylvania, Department of Mechanical Engineering and Applied Mechanics, Philadelphia, PA 19104, USA
[4] École Polytechnique, Département de Mécanique, 91128 Palaiseau cedex, France

* Corresponding author. *E-mail:* groger@lanl.gov

## Abstract

Based on the atomistic studies presented in Part I we develop analytical yield criteria for single crystals that capture the effect of shear stresses other than the Schmid stress (non-glide stresses) on the shear stress needed for dislocation glide (Peierls stress). These yield criteria characterize a non-associated plastic flow that originates owing to the complex response of 1/2⟨111⟩ screw dislocations to an applied stress tensor. Employing these criteria we identify the operative slip systems for tensile/compressive loading along various axes within the standard stereographic triangle and determine the ensuing tension-compression asymmetry. This result is in an excellent qualitative agreement with available experimental data. Moreover, using the constructed yield criteria within the Taylor homogenization procedure, we demonstrate that effects associated with non-planar cores of screw dislocations persist in random polycrystals. This affects significantly critical phenomena such as shear localization, which is demonstrated by analyzing the cavitation in a ductile plastic solid.

*Keywords:* non-glide stresses; yield criterion; non-associated flow; yield surface; strength differential; random polycrystal; cavitation instability

## 1. Introduction

Plastic deformation of pure body-centered cubic (BCC) metals is primarily controlled by the glide of 1/2⟨111⟩ screw dislocations that have high lattice friction (Peierls) stress owing to their non-planar cores (for reviews see [1-9]). In the Part I of this series we presented the results of atomistic studies of the core structure and glide of 1/2⟨111⟩ screw dislocations at 0 K in molybdenum and tungsten that employed the recently developed Bond Order Potentials [10, 11]. The principal conclusion is that the Peierls stress, identified with the critical resolved shear (CRSS) that acts parallel to the slip direction in the maximum resolved shear stress plane (MRSSP) at which the dislocation starts to move, is a function of both the orientation of the MRSSP and the magnitude of the shear stress perpendicular to the slip direction. The orientation of the MRSSP is determined by the angle $\chi$, defined in Fig. 2 of Part I, and the shear stress perpendicular to the slip direction, $\tau$, defined by equation (2) in Part I. The glide plane is always of the {110} type although not necessarily the {110} plane with the highest shear stress in the slip direction. Comparison with earlier atomistic studies that employed central-force potentials [2, 3, 5, 12-16] suggests that this conclusion is general and applies qualitatively to all BCC metals and possibly also to other materials with this crystal structure. However, quantitatively the dependence of the CRSS on $\chi$ and $\tau$ differs rather significantly from material to material; this is visible in Figs. 4, 7 and 8 of Part I that display such dependencies for molybdenum and tungsten, respectively.

The calculations presented in Part I were all done for the slip direction [111] and the angle $\chi$ was measured from the $(\bar{1}01)$ plane. Owing to the three-fold screw axis symmetry of the [111] direction the results would be the same if we chose to measure $\chi$ from any of the other two {110} planes of the [111] zone, $(01\bar{1})$ or $(1\bar{1}0)$. Furthermore, in any BCC crystal there are twelve independent systems comprised of $\langle 111 \rangle$ slip directions and {110} slip planes. They are all crystallographically equivalent and thus the results found in Part I apply equally to all of them. Hence, when we investigate the plastic yielding of a BCC single crystal all these systems have to be considered. Such analysis for loading by a general stress tensor is presented in Section 2 and a specific example of a combined loading by shear stresses parallel and perpendicular to the [111] direction is the topic of Section 3. In the latter the dependencies of the CRSS on $\chi$ and $\tau$, determined in Part I, have been employed. These numerically evaluated dependencies could, of course, be used when analyzing any type of loading. However, it is much more efficient and astute to formulate an analytical yield criterion that applies to any $\{110\}\langle 111 \rangle$ system and reproduces with sufficient accuracy the numerical results. This approach is analogous to that first advanced by Taylor [17] and later developed to include capabilities that account for finite shape change and lattice rotations and applied in studies of multislip hardening and strain localization [18-26].

These continuum models all assume the Schmid-type plastic behavior, which means that the only stress component that affects the yielding is the shear stress acting in the slip plane parallel to the slip direction [27, 28]. This is well-established for slip confined to {111} planes in FCC metals and basal planes in HCP metals. Since the plastic flow mediated by dislocations is always driven by the shear stress in the direction of slip, the plastic flow obeying the Schmid law is called associated. However, if also other stress components, called non-glide stresses, affect the yielding and plastic flow but they do not drive the dislocation glide in the slip plane, the Schmid law does not apply and such flow is called non-associated [26, 29]; more exact definition is presented in Section 8. The latter applies in BCC metals and the primary objective of this paper is to formulate the yield criterion for the non-associated flow that includes both the shear stresses parallel and perpendicular to the slip direction that affect the dislocation glide. This is done following the suggestion by Qin and Bassani [29, 30] made when formulating the yield criterion for $L1_2$ intermetallic compounds deformed in the anomalous regime of the temperature dependence of the yield stress [31]. The yield criterion formulated in Section 4 of this paper, which reproduces both the $\chi$ and $\tau$ dependence of the CRSS, is written as a linear combination of four shear stresses, the Schmid stress and three non-glide stresses, one parallel and two perpendicular to the slip direction. This approach was employed earlier in the special case of loading by pure shear stress parallel to the slip direction [32-34].

As the next step we employ the constructed yield criterion to determine the yield surface and compare this to a hypothetical yield surface that would be obtained if the Schmid law were valid. Furthermore, we use the yield criterion to determine active slip systems that operate during uniaxial loading for all orientations of the loading axis within the standard stereographic triangle. This analysis reveals a very notable contrast between tension and compression as well as a significantly different slip geometry in molybdenum and tungsten. Results of this study are then compared with recent experimental investigation of tension-compression asymmetries in molybdenum [35-37]. The remarkable agreement between the present calculations and experimental data demonstrates that the tension-compression asymmetry, commonly associated with the twinning-antitwinning asymmetry of shearing [1], is only partly related to this crystallographic characteristic. The major contribution originates in the effect of shear stresses perpendicular to the slip direction. In Section 8 we utilize the Taylor homogenization model to demonstrate that this asymmetry persists even in random polycrystals and in Section 9 we show on the example of ductile cavitation that the non-glide stresses also significantly affect critical plastic phenomena.

## 2. Plastic flow in a BCC single crystal subject to external loading

The atomistic calculations presented in the Part I of this series of papers were all done for the 1/2[111] screw dislocation. The angle $\chi$, determining the orientation of the MRSSP in which the shear stress $\sigma$ parallel to the [111] direction is applied, was measured from the $(\bar{1}01)$ plane. Owing to the crystal symmetry only $-30^\circ \leq \chi \leq +30^\circ$ need to be considered. The shear stress $\tau$ perpendicular to the slip direction is applied by Eq. (2) of Part I. Hence, the $(\bar{1}01)$ plane is the {110} plane with the highest shear stress parallel to the slip direction in between the three {110} planes of the [111] zone and the $(\bar{1}01)$ [111] system was chosen as a *reference system* in the atomistic studies presented in Part I. However, in any BCC crystal there are twelve crystallographically equivalent $\{110\}\langle 111\rangle$ systems and thus the dependencies of the CRSS on $\chi$ and $\tau$, found in Part I, apply equally to all of them. In addition, for $\chi \neq 0$ the CRSS depends on the sense of shearing and changing the sign of $\sigma$ is equivalent to changing the sign of $\chi$ while keeping the sign of $\sigma$ fixed. This corresponds to the so-called twinning-antitwinning asymmetry and an alternative way to capture this effect is to regard positive and negative slip directions as distinct [1, 2]. In this case only positive $\sigma$, and thus positive CRSS, need to be considered. The ensuing twenty four $\{110\}\langle 111\rangle$ reference systems are summarized in Table 1 and denoted by $\alpha$; the system used in atomistic studies corresponds to $\alpha = 2$. All these systems have to be included on equal footing when analyzing the plastic response of a single crystal to the loading represented by an applied stress tensor.

In such analysis we have to find first the orientations of the MRSSPs for the eight distinct slip directions and determine the shear stresses parallel and perpendicular to these slip directions in the MRSSPs associated with them. As explained above, only positive shear stresses parallel to the slip directions are considered and thus we exclude all reference systems for which these stresses are negative. This means that at this point only four slip directions remain in the subsequent analysis. In the next step we find for each of these four slip directions the {110} plane of its zone that has the highest shear stress parallel to the slip direction. Combination of this {110} plane with the concurrent slip direction represents a reference system that is always one of the systems listed in Table 1. The orientation of the MRSSP is for a system $\alpha$ characterized by the angle $\chi_\alpha$ that it makes with the corresponding {110} reference plane. The range of this angle is $-30^\circ \leq \chi_\alpha \leq +30^\circ$ and it is bounded by two {112} planes. For $\chi_\alpha < 0$ the nearest {112} plane is sheared in the *twinning* sense while for $\chi_\alpha > 0$ the nearest {112} plane is sheared in the *antitwinning* sense. This definition conforms to that used for the reference system $(\bar{1}01)[111]$ ($\alpha = 2$) in Part I.

Table 1: The 24 slip systems in bcc crystals. Note, that the crystallographic vectors $\mathbf{m}^{\alpha}$, $\mathbf{n}^{\alpha}$, $\mathbf{n}_1{}^{\alpha}$ have to be normalized before their use in (5).

| $\alpha$ | ref. system | $\mathbf{m}^{\alpha}$ | $\mathbf{n}^{\alpha}$ | $\mathbf{n}_1{}^{\alpha}$ | $\alpha$ | ref. system | $\mathbf{m}^{\alpha}$ | $\mathbf{n}^{\alpha}$ | $\mathbf{n}_1{}^{\alpha}$ |
|---|---|---|---|---|---|---|---|---|---|
| 1 | $(01\bar{1})[111]$ | $[111]$ | $[01\bar{1}]$ | $[\bar{1}10]$ | 13 | $(01\bar{1})[\bar{1}\bar{1}\bar{1}]$ | $[\bar{1}\bar{1}\bar{1}]$ | $[01\bar{1}]$ | $[10\bar{1}]$ |
| 2 | $(\bar{1}01)[111]$ | $[111]$ | $[\bar{1}01]$ | $[0\bar{1}1]$ | 14 | $(\bar{1}01)[\bar{1}\bar{1}\bar{1}]$ | $[\bar{1}\bar{1}\bar{1}]$ | $[\bar{1}01]$ | $[\bar{1}10]$ |
| 3 | $(1\bar{1}0)[111]$ | $[111]$ | $[1\bar{1}0]$ | $[10\bar{1}]$ | 15 | $(1\bar{1}0)[\bar{1}\bar{1}\bar{1}]$ | $[\bar{1}\bar{1}\bar{1}]$ | $[1\bar{1}0]$ | $[0\bar{1}1]$ |
| 4 | $(\bar{1}0\bar{1})[\bar{1}11]$ | $[\bar{1}11]$ | $[\bar{1}0\bar{1}]$ | $[\bar{1}\bar{1}0]$ | 16 | $(\bar{1}0\bar{1})[1\bar{1}\bar{1}]$ | $[1\bar{1}\bar{1}]$ | $[\bar{1}0\bar{1}]$ | $[01\bar{1}]$ |
| 5 | $(0\bar{1}1)[\bar{1}11]$ | $[\bar{1}11]$ | $[0\bar{1}1]$ | $[101]$ | 17 | $(0\bar{1}1)[1\bar{1}\bar{1}]$ | $[1\bar{1}\bar{1}]$ | $[0\bar{1}1]$ | $[\bar{1}\bar{1}0]$ |
| 6 | $(110)[\bar{1}11]$ | $[\bar{1}11]$ | $[110]$ | $[01\bar{1}]$ | 18 | $(110)[1\bar{1}\bar{1}]$ | $[1\bar{1}\bar{1}]$ | $[110]$ | $[101]$ |
| 7 | $(0\bar{1}\bar{1})[\bar{1}\bar{1}1]$ | $[\bar{1}\bar{1}1]$ | $[0\bar{1}\bar{1}]$ | $[1\bar{1}0]$ | 19 | $(0\bar{1}\bar{1})[11\bar{1}]$ | $[11\bar{1}]$ | $[0\bar{1}\bar{1}]$ | $[\bar{1}0\bar{1}]$ |
| 8 | $(101)[\bar{1}\bar{1}1]$ | $[\bar{1}\bar{1}1]$ | $[101]$ | $[011]$ | 20 | $(101)[11\bar{1}]$ | $[11\bar{1}]$ | $[101]$ | $[1\bar{1}0]$ |
| 9 | $(\bar{1}10)[\bar{1}\bar{1}1]$ | $[\bar{1}\bar{1}1]$ | $[\bar{1}10]$ | $[\bar{1}0\bar{1}]$ | 21 | $(\bar{1}10)[11\bar{1}]$ | $[11\bar{1}]$ | $[\bar{1}10]$ | $[011]$ |
| 10 | $(10\bar{1})[1\bar{1}1]$ | $[1\bar{1}1]$ | $[10\bar{1}]$ | $[110]$ | 22 | $(10\bar{1})[\bar{1}1\bar{1}]$ | $[\bar{1}1\bar{1}]$ | $[10\bar{1}]$ | $[0\bar{1}\bar{1}]$ |
| 11 | $(011)[1\bar{1}1]$ | $[1\bar{1}1]$ | $[011]$ | $[\bar{1}01]$ | 23 | $(011)[\bar{1}1\bar{1}]$ | $[\bar{1}1\bar{1}]$ | $[011]$ | $[110]$ |
| 12 | $(\bar{1}\bar{1}0)[1\bar{1}1]$ | $[1\bar{1}1]$ | $[\bar{1}\bar{1}0]$ | $[0\bar{1}\bar{1}]$ | 24 | $(\bar{1}\bar{1}0)[\bar{1}1\bar{1}]$ | $[\bar{1}1\bar{1}]$ | $[\bar{1}\bar{1}0]$ | $[\bar{1}01]$ |

Finally, for each reference system included we transform the externally applied stress tensor into the right-handed coordinate system with the *z*-axis parallel to the corresponding $\langle 111 \rangle$ direction, *y*-axis normal to the MRSSP and the *x*-axis in the MRSSP. This is congruent with the coordinate system defined for the reference system $(\bar{1}01)[111]$ in conjunction with the application of $\tau$ by Eq. 2 in Part I. In general, for this orientation all components of the applied stress tensor are nonzero. However, as shown in Part I, the only stress components affecting the glide of screw dislocations are the resolved shear stresses $\sigma_\alpha$ and $\tau_\alpha$, parallel and perpendicular to the slip direction of the reference system $\alpha$, respectively. Hence, we extract from the full applied stress tensor the tensor

$$\mathbf{\Sigma}^{\alpha}(\chi_\alpha) = \begin{bmatrix} -\tau_\alpha & 0 & 0 \\ 0 & \tau_\alpha & \sigma_\alpha \\ 0 & \sigma_\alpha & 0 \end{bmatrix}. \qquad (1)$$

This tensor determines for the reference system $\alpha$ the values of $\sigma_\alpha$ and $\tau_\alpha$ that arise for a given mode of loading when the MRSSP makes the angle $\chi_\alpha$ with the $\{110\}$ reference plane of the system $\alpha$. As the applied loading evolves, the shear stresses $\sigma_\alpha$ and $\tau_\alpha$ develop accordingly and the tensor $\mathbf{\Sigma}^{\alpha}(\chi_\alpha)$ defines a unique dependence of $\tau_\alpha$ on $\sigma_\alpha$. In the following we call their dependence the loading path $\sigma_\alpha - \tau_\alpha$.

The atomistic studies of the glide of the 1/2[111] screw dislocation suggest that for a given $\chi$ there is a unique relation between the CRSS and $\tau$, independent of the manner (i.e. loading history) the corresponding shear stresses $\sigma$ and $\tau$ were attained. For several values of $\chi$ this CRSS vs $\tau$ dependence was established in the studies presented in Part I. We then consider that a reference system $\alpha$ becomes activated for slip when $\sigma_\alpha$ that varies along a loading path $\sigma_\alpha - \tau_\alpha$ reaches the CRSS corresponding to the actual value of $\tau_\alpha$. In order to demonstrate such transfer of the atomistic results to the analysis of the plastic deformation of a single crystal, we discuss in the following section a combined loading by the shear stresses parallel and perpendicular to the slip direction. The same method will be employed subsequently to analyze uniaxial loading of single crystals.

## 3. Plastic yielding in a single crystal loaded by combined shear stresses parallel and perpendicular to the slip direction

In this section we consider that the applied stress tensor comprises a combination of pure shear stresses $\sigma$ and $\tau$, parallel and perpendicular to the slip direction. In this case the loading path along which a given combination of $\sigma$ and $\tau$ is attained is not unique. In fact it can be achieved in a limitless number of ways[1]. However, since the CRSS depends only on the value of $\tau$ and not on the way the corresponding combination of $\sigma$ and $\tau$ was produced, we will utilize paths specified by constant ratios of $\tau / \sigma$. For a given loading path the procedure outlined in the previous section determines the four reference systems $\alpha$ and the related pairs of stresses $\tau_\alpha$ and $CRSS_\alpha$ for which the plastic yielding would commence on these systems. The system with the lowest $CRSS_\alpha$ is the primary system on which the plastic deformation starts first.

We now consider the shear stress parallel to the [111] direction applied such that the $(\bar{1}01)$ plane is the MRSSP ($\chi$ = 0), combined with the applied shear stress perpendicular to this direction; in this case the reference system is $\alpha = 2$. This is the same loading as that considered in Section 4.3 of Part I for the 1/2[111] screw dislocation and the corresponding CRSS vs $\tau$ dependence is shown for Mo and W in Fig. 7 of Part I. Obviously, in the proximity of $\tau = 0$ the $(\bar{1}01)[111]$ ($\alpha$ = 2) is the primary slip system. However, as the magnitude of $\tau$ increases the shear stresses, both parallel and perpendicular to the slip direction, evolve also in other slip systems and which of the available systems is primary has to be investigated as described in the previous section. An example corresponding to the loading path $\tau_2 / \sigma_2 = 1.5$ for the system $\alpha = 2$ is shown in Fig. 1a where the squares are the data for molybdenum copied from Fig. 7a of Part I and represent the dependence of the CRSS on $\tau$ for the loading considered; the straight line passing through the origin depicts the loading path. If only 1/2[111] dislocations were gliding the $(\bar{1}01)[111]$ system would become operative at the point where this line intersects the CRSS vs $\tau$ dependence, i.e. at the point marked **B**. However, while the loading in the $(\bar{1}01)[111]$ reference system proceeds along the path shown in Fig. 1a another reference system, $(0\bar{1}1)[\bar{1}11]$ ($\alpha$ = 5) in which the orientation of the MRSSP corresponds to $\chi_5 = -7.3^\circ$, is subjected to the shear stresses $\sigma_5$ and $\tau_5$ that evolve along the loading path $\tau_5 / \sigma_5 \approx -0.2$. This path is shown as a straight line passing through the origin in Fig. 1b. It was determined employing the procedure outlined in the previous section using the stress tensor corresponding to the loading path of Fig. 1a. Note that the slopes of the two loading paths for the two reference systems are different. As already emphasized, the CRSS vs $\tau$ dependence is the same for every system $\alpha$ and thus the squares in Fig. 1b have the same meaning as in Fig. 1a but were now obtained for the MRSSP with angle $\chi = -9^{\circ}$ that is closest to $\chi_5$ in between the angles $\chi$ for which the CRSS vs $\tau$ dependence was calculated atomistically in Part I[2]. The system $(0\bar{1}1)[\bar{1}11]$ becomes operative at the point where the loading path in Fig. 1b intersects the CRSS vs $\tau$ dependence, i.e. at the point marked **A** in this figure. However, in the $(\bar{1}01)[111]$ system this stress corresponds to the loading at the point marked **A** in Fig. 1a, where the corresponding $\sigma_2$ is well below the CRSS. Consequently, the $(0\bar{1}1)[\bar{1}11]$ system will become operative *before* the $(\bar{1}01)[111]$ system and it is thus the primary system for the loading path considered. The same analysis as described above for molybdenum can also be performed for tungsten. The only difference is that in this case one uses the CRSS vs $\tau$ dependencies calculated for tungsten [38].

[1] In contrast, the loading path is unique in the usual experimental situations such as loading by tension/compression when it is linear.

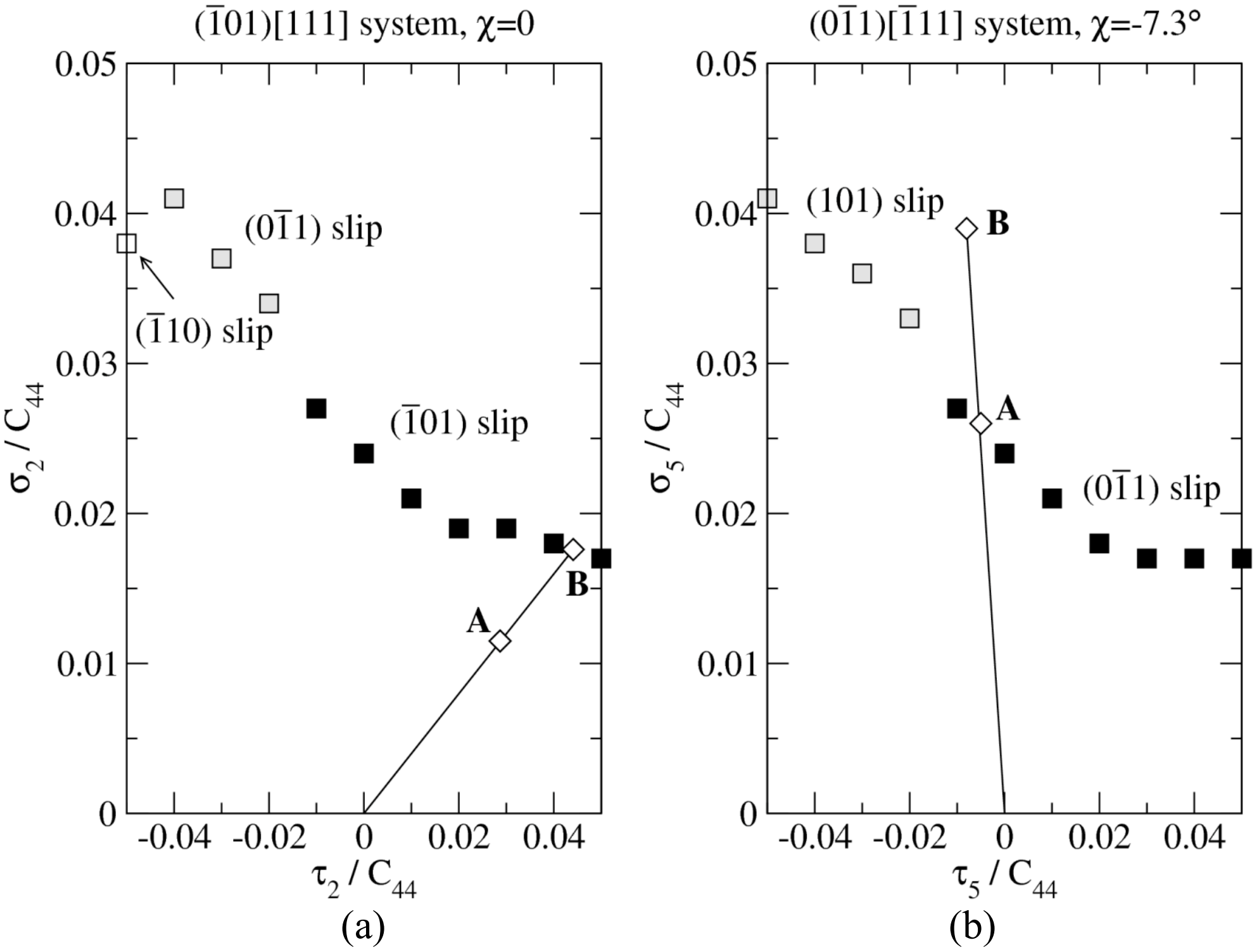

Fig. 1: Evolution of loading in two different $\{110\}\langle 111\rangle$ systems, represented by straight lines passing through the origin, induced by the shear stresses parallel and perpendicular to the slip direction applied in the $(\bar{1}01)[111]$ system: (a) $\alpha = 2$, $\tau_2 / \sigma_2 = 2.5$, (b) $\alpha = 5$, $\tau_5 / \sigma_5 = -0.2$. Squares correspond to the atomistic data obtained for an isolated 1/2[111] dislocation in Part I. The points marked **A** and **B** correspond to the same applied loading in both cases.

[2] This dependence was not shown in Part I but can be found in [38].

In order to develop a complete description of the yielding of a single crystal on the basis of the atomistic study of the glide of 1/2⟨111⟩ screw dislocations the procedure described above has to be repeated for a number of different loading paths $\tau_2 / \sigma_2 = \eta_2$ where $-\infty < \eta_2 < +\infty$ . For each of these loading paths such calculations yield four reference systems $\alpha$, each associated with a distinct slip direction, and for every system $\alpha$ a pair of stresses $CRSS_\alpha$ and $\tau_\alpha$ at which the plastic deformation commences on this system can be determined. It is seen from the results of Part I that for $\left|\tau_\alpha / C_{44}\right| \leq 0.02$ the dependence of $CRSS_\alpha$ on $\tau_\alpha$ is close to linear (see Figs. 7 and 8 of Part I). This is the range of $\tau_\alpha$ in which $CRSS_\alpha$ is always larger than $\tau_\alpha$ and in the following we consider only this regime of applied stresses[3]. In this case the $CRSS_\alpha$ evaluated for various loading paths follows in the plot of CRSS vs $\tau$ a straight line. For molybdenum such lines are drawn in Fig. 2a for the reference systems $\alpha$ = 5, 19, 2, 21, 8 and for tungsten in Fig. 2b for the reference systems $\alpha$ = 19, 2, 13, 17. The inner envelopes of these lines, drawn in Figs. 2a, b as the solid polygons, encompass the regions of the elastic behavior of the material. Hence, they represent projections of the yield surfaces (defined more precisely in Section 5) onto the CRSS vs $\tau$ plot. For any loading path characterized by a ratio $\tau_2 / \sigma_2$ the yielding occurs when this path reaches the yield surface, i.e. the corresponding polygon in Fig. 2. The {110}⟨111⟩ system associated with the side of the polygon that is intersected by the loading path is the primary slip system for this combination of $\sigma_2$ and $\tau_2$ .

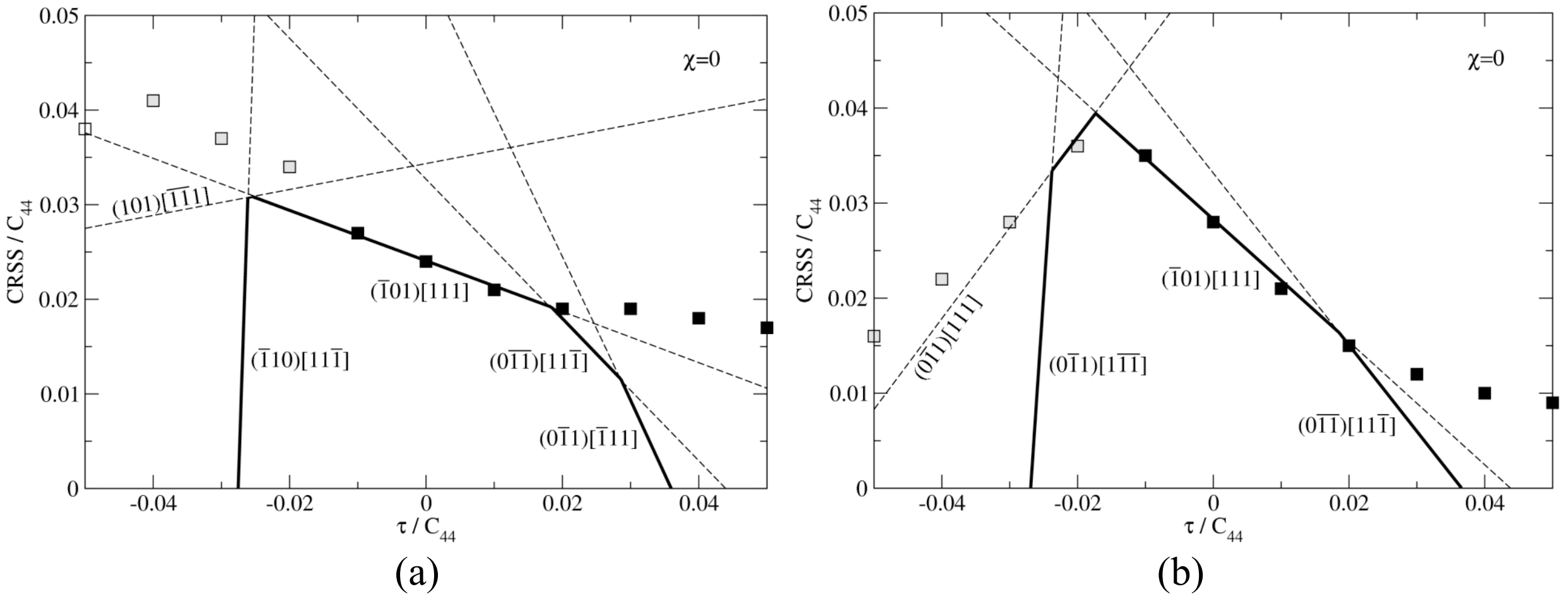

Fig. 2: Critical lines marking the onset of slip on slip systems in molybdenum (a) and tungsten (b), projected into the CRSS-$\tau$ plot for the MRSSP $(\bar{1}01)$ ($\chi$ = 0). The inner envelope of these lines (bold) defines a yield polygon for real single crystals. Squares correspond to the atomistic data obtained for an isolated 1/2[111] dislocation in Part I.

[3] The problem becomes more complex when $\left|\tau / C_{44}\right| > 0.02$ because the CRSS then depends on $\tau_\alpha$ non-linearly. For negative $\tau$ from this region the slip may occur on a {110} plane that is inclined with respect to the reference {110} plane by $\pm 60^\circ$. As discussed in Part I, this may be the origin of the anomalous slip that occurs at very low temperatures and takes place on the slip systems for which the resolved shear stress is substantially lower than that on the most highly stressed {110}⟨111⟩ system [39-51].

In Fig. 2b we show the projection of the yield surface calculated for tungsten using the procedure outlined above. Since the dependence of the CRSS on $\tau$ is now different from that for molybdenum, also the $\{110\}\langle 111\rangle$ systems that comprise the yield polygon are distinct from those in Fig. 2a for molybdenum.

## 4. Analytical yield criterion for single crystals

The dependence of the CRSS on the angle $\chi$ and the shear stress $\tau$ was in the previous section obtained directly from the results of atomistic calculations. However, for $|\tau / C_{44}| \leq 0.02$, i.e. when for a given $\chi$ the CRSS depends to a good approximation linearly on $\tau$, we can formulate an analytical yield criterion that reproduces very closely the atomistic data. In developing such criterion we follow the proposal of Qin and Bassani [29, 30] who introduced the yield criterion for non-associated flow in $Ni_3Al$ that captured the observed orientation dependence and tension-compression asymmetry of the yield stress in the regime of the anomalous increase of the flow stress with temperature (see e.g. [31]). This criterion included two shear stresses, one parallel and one perpendicular to the slip direction. In our case, in order to capture both the $\chi$ and $\tau$ dependence of the CRSS, such criterion needs to comprise two shear stresses parallel and two shear stresses perpendicular to the slip direction, both resolved in two different $\{110\}$ planes of the zone of the slip direction. For the [111] slip direction and reference system $\alpha = 2$ from Table 1, which was employed in the atomistic studies in Part I, we write such yield criterion as

$$\sigma^{(\bar{1}01)} + a_1\sigma^{(0\bar{1}1)} + a_2\tau^{(\bar{1}01)} + a_3\tau^{(0\bar{1}1)} = \tau^*_{cr} , \qquad (2)$$

where $\sigma^{\{110\}}$ and $\tau^{\{110\}}$ are shear stresses parallel and perpendicular to the slip direction, respectively, in the corresponding $\{110\}$ planes. The first term in Eq. (2) is the stress that drives the dislocation motion in the $(\bar{1}01)$ glide plane and does the work through dislocation glide. It is commonly called the Schmid stress. In contrast, the stresses $\sigma^{(0\bar{1}1)}$, $\tau^{(\bar{1}01)}$ and $\tau^{(0\bar{1}1)}$ affect the dislocation core but do not do any work when the dislocation glides in the $(\bar{1}01)$ plane. These stresses are commonly called non-glide stresses. The second term includes the shear stress parallel to the slip direction in another $\{110\}$ plane and reproduces the twinning-antitwinning asymmetry of the CRSS. For such loading a yield criterion employing just these two terms was developed earlier [32-34]. The third and fourth terms incorporate the effect of the shear stress perpendicular to the slip direction. The coefficients $a_1$, $a_2$, $a_3$ , as well as $\tau^*_{cr}$, are all adjustable parameters that are ascertained by fitting the CRSS vs $\chi$ and CRSS vs $\tau$ dependencies determined by atomistic calculations.

First, $a_1$ and $\tau^*_{cr}$ are fitted to reproduce the CRSS vs $\chi$ dependence for loading by pure shear stress parallel to the slip direction. In this case Eq. (2) reduces to $\sigma^{(\bar{1}01)} + a_1\sigma^{(0\bar{1}1)} = \tau^*_{cr}$. The two shear stresses entering this relation can be written in terms of the CRSS applied in the MRSSP and the angle $\chi$ as $\sigma^{(\bar{1}01)} = \mathrm{CRSS}\cos\chi$ and $\sigma^{(0\bar{1}1)} = \mathrm{CRSS}\cos(\chi + \pi/3)$. For a given orientation of the MRSSP, i.e. angle $\chi$, the corresponding CRSS can be determined by the criterion (2) as

$$\mathrm{CRSS}(\chi) = \frac{\tau^*_{cr}}{\cos\chi + a_1\cos(\chi + \pi/3)} . \qquad (3)$$

$a_1$ and $\tau^*_{cr}$ can then be ascertained by the least squares fitting of this relation to the CRSS vs $\chi$ dependence found in atomistic studies (Fig. 4 of Part I). For molybdenum $a_1$ and $\tau^*_{cr}$ were found in this way in [32-34] and it can be seen in these papers that Eq. (3) reproduces the atomistically determined CRSS vs $\chi$ dependence very closely for all values of $\chi$.

In the second step, keeping $a_1$ and $\tau^*_{cr}$ fixed, the parameters $a_2$ and $a_3$ have been determined by fitting the CRSS vs $\tau$ dependence found in the atomistic calculations that employed the combination of the shear stresses perpendicular and parallel to the slip direction (Section 4.3 of Part I). In this case $\tau^{(\bar{1}0\bar{1})} = \tau \sin 2\chi$ and $\tau^{(0\bar{1}\bar{1})} = \tau \cos(2\chi + \pi/6)$. For a given angle $\chi$ and shear stress $\tau$ the CRSS can be determined by the criterion (2) such that

$$\mathrm{CRSS}(\chi,\tau) = \frac{\tau^*_{cr} - \tau[a_2 \sin 2\chi + a_3 \cos(2\chi + \pi/6)]}{\cos\chi + a_1 \cos(\chi + \pi/3)} . \tag{4}$$

The coefficients $a_2$ and $a_3$ can again be ascertained by the least squares fitting of this relation to the CRSS vs $\tau$ dependencies calculated for various angles $\chi$. For $|\tau / C_{44}| \leq 0.02$, which is the range of $\tau$ for which the yield criterion has been developed, we found that the best fit is obtained when considering only three orientations of the MRSSP, namely $\chi = 0$ and $\chi \approx \pm 9^\circ$. In addition, for each of these orientations, only the data for $\tau/C_{44} = \pm 0.01$ need to be taken into account since only two points specify the slope of the straight line that approximates the CRSS vs $\tau$ dependence in the region of $\tau$ specified above. The coefficients $a_1$, $a_2$, $a_3$ and $\tau^*_{cr}$ entering the yield criterion (2) for molybdenum and tungsten, which were determined as described above, are listed in Table 2.

Table 2: Coefficients in the yield criterion (5) for molybdenum and tungsten determined by fitting the results of 0 K atomistic studies.

| | $a_1$ | $a_2$ | $a_3$ | $\tau^*_{\mathrm{cr}} / C_{44}$ |
|---|---|---|---|---|
| molybdenum | 0.24 | 0 | 0.35 | 0.027 |
| tungsten | 0 | 0.56 | 0.75 | 0.028 |

The yield criterion (2) applies, of course, to any $\{110\}\langle 111\rangle$ reference system. For a general loading by an applied stress tensor $\mathbf{\Sigma}^{app}$ defined in the cube axes this criterion can be written for any system $\alpha$ listed in Table 1 as

$$\mathbf{m}^\alpha \mathbf{\Sigma}^{app} \mathbf{n}^\alpha + a_1 \mathbf{m}^\alpha \mathbf{\Sigma}^{app} \mathbf{n}_1^\alpha + a_2 (\mathbf{n}^\alpha \times \mathbf{m}^\alpha) \mathbf{\Sigma}^{app} \mathbf{n}^\alpha + a_3 (\mathbf{n}_1^\alpha \times \mathbf{m}^\alpha) \mathbf{\Sigma}^{app} \mathbf{n}_1^\alpha = \tau^*_{cr} , \tag{5}$$

where $\mathbf{m}^\alpha$ is the unit vector in the slip direction, $\mathbf{n}^\alpha$ the unit vector perpendicular to the reference plane, and $\mathbf{n}_1^\alpha$ the unit vector perpendicular to the $\{110\}$ plane in the zone of $\mathbf{m}^\alpha$ that makes the angle $-60^\circ$ with the reference plane. This criterion determines a critical applied stress tensor, $\mathbf{\Sigma}_c^{app}$, at which the yielding on the system $\alpha$ commences. Recall that an angle is positive if measured relative to the reference plane in the sense shown in Fig. 2 of Part I. For example, if the reference system is $(\bar{1}01)[111]$, i.e. $\alpha = 2$, the three unit vectors are $\mathbf{m}^2 = \frac{1}{\sqrt{3}}[111]$, $\mathbf{n}^2 = \frac{1}{\sqrt{2}}[\bar{1}01]$ and $\mathbf{n}_1^2 = \frac{1}{\sqrt{2}}[0\bar{1}1]$. The complete list of the three vectors $\mathbf{m}^\alpha$, $\mathbf{n}^\alpha$ and $\mathbf{n}_1^\alpha$ for all twenty four $\{110\}\langle 111\rangle$ systems is given in Table 1. Notice that the reference systems labeled 1 to 12 and 13 to 24 are conjugate to each other in that the pairs of systems $\alpha$ and $\alpha$+12 have identical normals of the reference planes $\mathbf{n}^{\alpha}$, opposite slip directions $\mathbf{m}^{\alpha}$, and complementary vectors $\mathbf{n}_1^{\alpha}$.

In the criterion (5) the last three terms represent the effect of non-glide stresses and the first term is the Schmid stress acting in the slip plane of the system $\alpha$. For any applied loading given by the stress tensor $\mathbf{\Sigma}^{app}$, one can assess the activity of individual reference systems summarized in Table 1 by evaluating the left side of equation (5), which we mark $\tau^{*\alpha}$, for $\alpha$ ranging from 1 to 24. Following the criterion (5) the slip on a system $\alpha$ is considered activated when $\tau^{*\alpha}$ reaches $\tau^*_{cr}$ and thus when the applied stress tensor attains the critical value $\mathbf{\Sigma}^{app}_c$. In this framework the plastic deformation at 0 K starts when one of the 24 values of $\tau^{*\alpha}$ reaches $\tau^*_{cr}$. The corresponding $\alpha$ then identifies the slip system that will be activated first and, in the following, this reference system is called the primary system.

## 5. The yield surface

In order to introduce the notion of the yield surface we define the so-called principal space that is a three-dimensional space with orthogonal axes along which we measure the three principal stresses of any applied stress tensor. Hence, any stress state is represented in this space by a point the coordinates of which are the corresponding principal stresses $\sigma_I$, $\sigma_{II}$ and $\sigma_{III}$. The yield surface, determined by the yield criterion, is a convex hyperplane in the principal space that encompasses the region of stress states for which the deformation behavior of the material is purely elastic. When the stress state corresponding to an applied loading reaches the yield surface, the yield criterion is satisfied and the material starts to deform plastically.

For a given yield criterion the yield surface can be constructed as follows. In order to explore all possible orientations of the applied loading we write $\sigma_I = \kappa \sin\theta \cos\phi$, $\sigma_{II} = \kappa \sin\theta \sin\phi$ and $\sigma_{III} = \kappa \cos\theta$, where $\theta \in \langle 0,\pi \rangle$ and $\phi \in \langle 0,2\pi \rangle$ are spherical angles. Every combination of angles $\theta$ and $\phi$ corresponds to a well-defined direction in the principal space and for each such combination we evaluate $\sigma_I$, $\sigma_{II}$, $\sigma_{III}$ and determine $\kappa$ such that the yield criterion is satisfied. For this value of $\kappa$, which is a function of $\theta$ and $\phi$, the principal stresses determine the critical stress state for which the yielding commences and thus the point on the yield surface. When this calculation is repeated for all angles $\theta$ and $\phi$, one obtains the full three-dimensional yield surface. In the framework of crystal plasticity the procedure described above has to be repeated for every available slip system and the inner envelope of all calculated critical stress states then represents the yield surface.

When employing the yield criterion (5) we first calculate for every combination of angles $\theta$ and $\phi$ the principal stresses $\sigma_I$, $\sigma_{II}$, $\sigma_{III}$ and thus the stress tensor $\mathbf{\Sigma}^{app}(\theta,\phi)$. We then evaluate for all slip systems $\alpha$ from Table 1 the left side of Eq. (5) using the tensor $\mathbf{\Sigma}^{app}(\theta,\phi)$. If we mark this value $\tau^{*\alpha}(\theta,\phi)$ then for these angles $\theta$ and $\phi$ the yielding will commence on the slip system $\alpha$ when $\kappa$ is equal to $\kappa^{\alpha}_c(\theta,\phi) = \tau^*_{cr} / \tau^{*\alpha}(\theta,\phi)$. For the slip system $\alpha$ the stresses $\sigma^{\alpha}_I = \kappa^{\alpha}_c(\theta,\phi)\sin\theta\cos\phi$, $\sigma^{\alpha}_{\mathrm{II}} = \kappa^{\alpha}_{\mathrm{c}}(\theta,\phi)\sin\theta\sin\phi$ and $\sigma^{\alpha}_{III} = \kappa^{\alpha}(\theta,\phi)\cos\theta$ determine the critical stress state for which this system becomes activated for slip. The yield surface is then the inner envelope of the critical stresses $\sigma^{\alpha}_I$, $\sigma^{\alpha}_{II}$, $\sigma^{\alpha}_{III}$ calculated for all angles $\theta$ and $\phi$ and all slip systems $\alpha$ from Table 1. In general, different slip systems operate in different parts of the yield surface.

Although the stress tensor $\mathbf{\Sigma}^{app}$ entering the yield criterion (5) may contain nonzero hydrostatic stress, this does not affect the magnitude of $\tau^{*\alpha}$ and thus the onset of yielding. Hence, it is a common practice to represent the yield surface as a two-dimensional yield locus that is obtained as a cross-section of the three-dimensional yield surface by the so-called deviatoric plane whose normal is parallel to the direction in which $\sigma_I = \sigma_{II} = \sigma_{III}$ [52]. The three-dimensional yield surface is then uniquely characterized by its yield locus in the deviatoric plane. The yield loci for the yield surfaces determined by the procedure described above are shown in Fig. 3 for molybdenum (dashed polygon) and for tungsten (solid polygon). For comparison, we also plot in this figure the yield locus that is predicted by the Schmid law (dotted polygon), i.e. when $a_1 = a_2 = a_3 = 0$ in (5). In this case, the yield criterion reduces to that of Tresca and the yield locus is a regular hexagon. The remarkable difference between the yield loci obtained from the yield criterion (5) and that corresponding to the Schmid law demonstrates the breakdown of the Schmid law in both molybdenum and tungsten. In molybdenum, this is caused by a combination of the twinning-antitwinning asymmetry of shearing parallel to the slip direction and the effect of the shear stresses perpendicular to the slip direction, while in tungsten only the latter plays role.

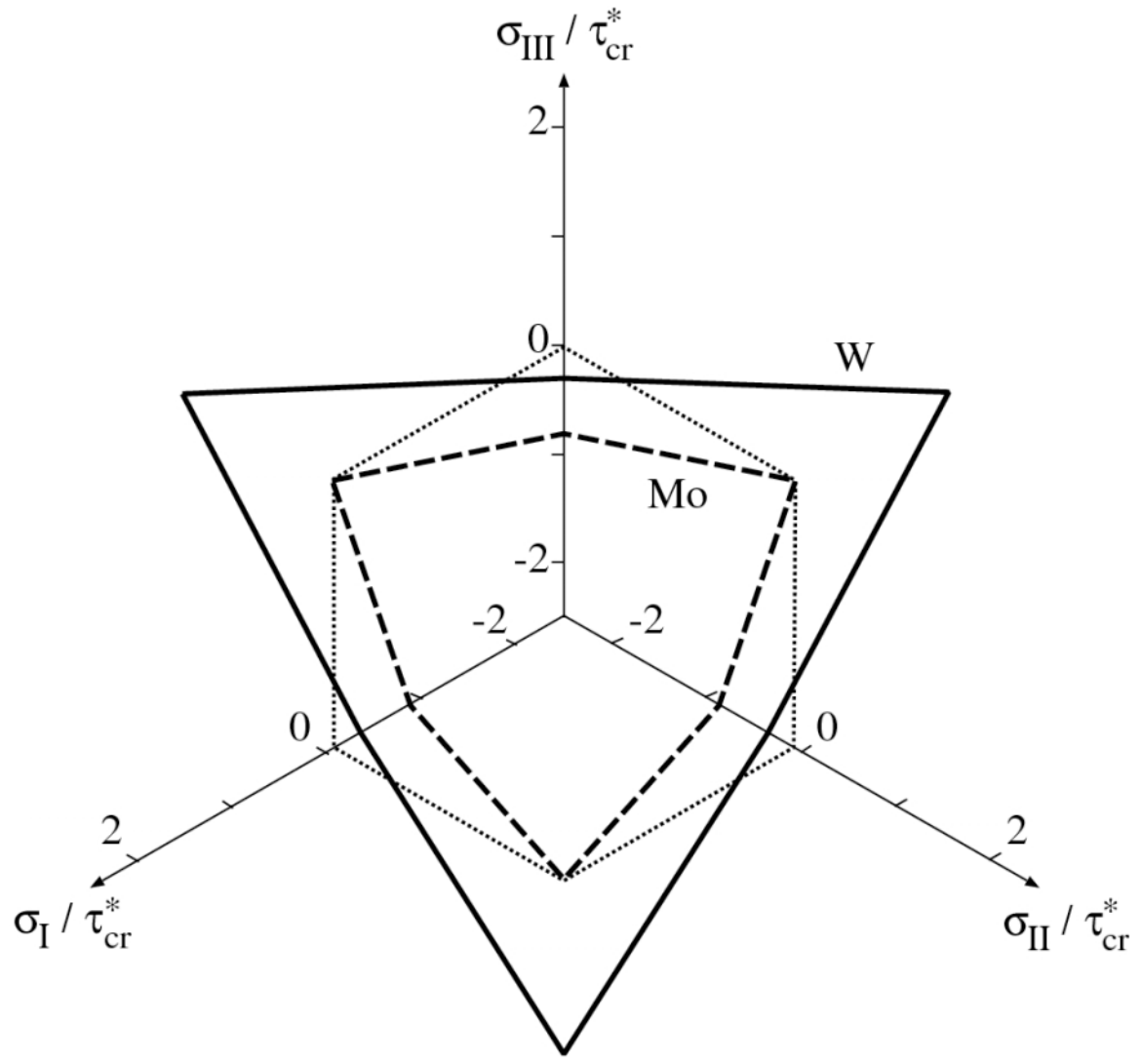

Fig. 3: The yield loci in the deviatoric plane generated by the yield criterion (5) for molybdenum (dashed polygon) and for tungsten (solid polygon). For comparison we also show as the dotted polygon the yield locus that is obtained from the Schmid law ($a_1$=$a_2$=$a_3$=0). In this projection, each edge of the yield locus is shared by four slip systems and the corners by eight slip systems.

## 6. Application of the yield criterion to uniaxial loading

In this section we discuss tensile and compressive loadings along the directions for which the most highly stressed $\{110\}\langle 111\rangle$ system is $(\bar{1}01)[111]$. In Section 4.2 of Part I several such cases were studied atomistically and orientations of the corresponding loading axes were summarized in Fig. 5 of Part I. Assuming a unit uniaxial applied stress (+1 for tension and –1 for compression) we evaluate first for each loading axis studied the corresponding applied stress tensor $\mathbf{\Sigma}^{app}$, defined in Section 4 in connection with equation (5). In order to consider all possible slip systems that can operate for a given loading axis we identify the four relevant reference systems employing the method described in Section 2. These systems are in between the 24 systems marked by $\alpha$ in Table 1. As the next step we determine for these systems the corresponding vectors $\mathbf{m}^{\alpha}$, $\mathbf{n}^{\alpha}$ and $\mathbf{n}_1^{\alpha}$, defined in Section 4. Finally, we evaluate the left side of Eq. (5) and mark it $\tau^{*\alpha}_{t/c}$. According to the yield criterion (5) the uniaxial tensile/compressive stress for which a system $\alpha$ becomes activated if $\sigma^{\alpha}_{t/c} = \tau^{*}_{cr} / \tau^{*\alpha}_{t/c}$. The actual *uniaxial* yield stress that induces plastic flow in the crystal corresponds to the minimum among these stresses, i.e. $\sigma_{t/c} = \min_{\alpha} \sigma^{\alpha}_{t/c}$, and the corresponding system $\alpha$ is then the primary system.

Following this procedure we find that in tension (positive $\tau$) the $(\bar{1}01)[111]$ system, which has the highest Schmid stress, is the primary slip system for any orientation of the loading axis within the stereographic triangle for both molybdenum and tungsten. However, in compression (negative $\tau$) the primary slip systems vary considerably with the orientation of the loading axis. Regions of different slip systems for different orientations of compressive axes within the stereographic triangle are depicted in Fig. 4a for molybdenum and in Fig. 4b for tungsten. Since the sense of shearing is reversed relative to tension and we consider that shear stresses parallel to the slip directions are always positive (see Section 2), the slip system with the highest Schmid stress is not $(\bar{1}01)[111]$ but its conjugate $(\bar{1}01)[\bar{1}\,\bar{1}\,\bar{1}]$.

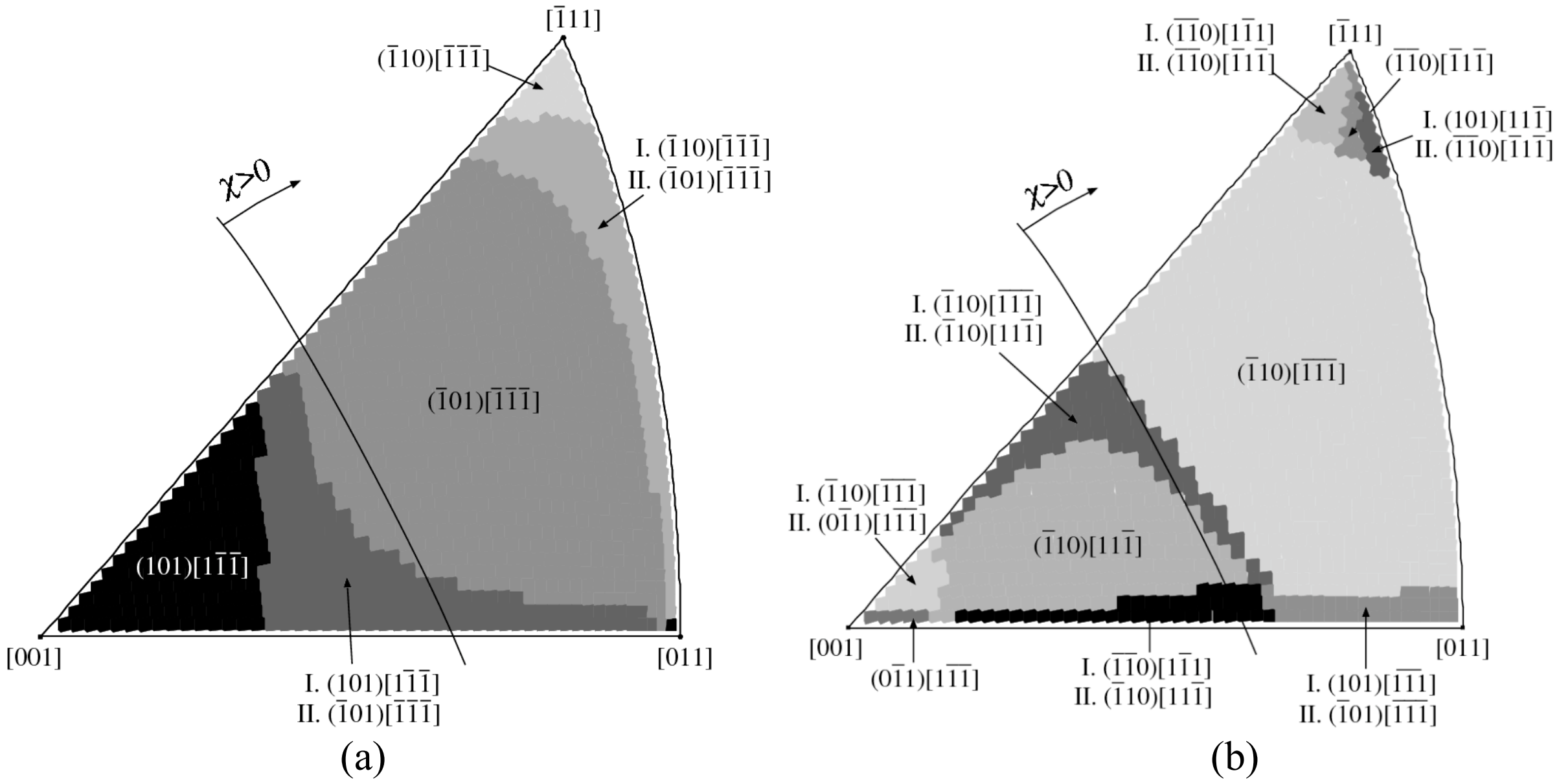

Fig. 4: Primary slip systems in *compression* predicted from the yield criterion (5) for molybdenum (a) and tungsten (b). In the regions of simultaneous activity of two slip systems, the primary slip system with lower yield stress is marked as I and the secondary, whose yield stress is at most 2% higher than I, as II.

Consider first the distribution of primary slip systems for molybdenum in Fig. 4a. In the central region of the triangle, the most operative slip system is, indeed, $(\bar{1}01)[\bar{1}\,\bar{1}\,\bar{1}]$. As the loading axis moves towards the $[011]-[\bar{1}11]$ edge, the $(\bar{1}10)[\bar{1}\,\bar{1}\,\bar{1}]$ system, which dominates near the $[\bar{1}11]$ corner, becomes increasingly more prominent. There is a region in which the two systems can operate simultaneously, i.e. the uniaxial yield stress of the secondary system (marked as II) is within 2% of that of the primary system (denoted as I). This could give rise to the macroscopic slip on the $(\bar{2}11)$ plane. A similar situation arises on the other side of the triangle. For orientations close to the [001] corner the slip system $(101)[1\bar{1}\bar{1}]$ dominates but there is again an intermediate region where the slip systems $(\bar{1}01)[\bar{1}\bar{1}\bar{1}]$ and $(101)[1\bar{1}\bar{1}]$ can be equally active. However, the slip directions are now different and involve 1/2[111] and $1/2[\bar{1}11]$ dislocations that move simultaneously. This so-called multislip mechanism of plastic deformation has been frequently observed in low-temperature experiments not only in molybdenum but also in other pure BCC refractory metals.

In the case of tungsten, the distribution of primary slip systems, shown in Fig. 4b, is rather different and more complex than for molybdenum. The striking difference between the two metals is that while in molybdenum the $(\bar{1}01)[\bar{1}\bar{1}\bar{1}]$ system dominates the central region of the stereographic triangle in which $\chi>0$, in tungsten it is the $(\bar{1}10)[\bar{1}\bar{1}\bar{1}]$ system that is dominant in this region. Only near the [011] corner the slip systems $(\bar{1}01)[\bar{1}\bar{1}\bar{1}]$ and $(\bar{1}10)[\bar{1}\bar{1}\bar{1}]$ may operate simultaneously. It is important to emphasize that the Schmid factor for the $(\bar{1}10)[\bar{1}\bar{1}\bar{1}]$ system is only a half of that for the $(\bar{1}01)[\bar{1}\bar{1}\bar{1}]$ system. The region of negative $\chi$ is dominated by the $(\bar{1}10)[11\bar{1}]$ system surrounded by regions of simultaneous operation of the $(\bar{1}10)[11\bar{1}]$ system and $(\bar{1}10)[\bar{1}\bar{1}\bar{1}]$ or $(\bar{1}\bar{1}0)[1\bar{1}1]$, respectively. Similarly as in molybdenum, this results in multislip deformation that takes place as a consequence of cooperative motion of two different $1/2\langle 111\rangle$ dislocations. As seen in Fig. 4b, a variety of slip systems, several of them leading to the multislip, are found near the [001] and $[\bar{1}11]$ corners of the stereographic triangle.

The reason for the significant difference between molybdenum and tungsten when loaded in compression is that in molybdenum the difference between orientations corresponding to positive and negative $\chi$ is to a great extent governed by the twinning-antitwinning asymmetry encountered in pure shear along the slip direction of the most highly stressed $(\bar{1}01)[\bar{1}\bar{1}\bar{1}]$ system. In contrast, as shown in Part I, this asymmetry plays virtually no role in tungsten and the choice of the most favored slip system is affected solely by the shear stress perpendicular to the slip direction.

Unfortunately experiments currently accessible are not sufficient to test the predicted variety of slip systems in molybdenum and tungsten loaded in tension/compression at very low temperatures. Especially, no low temperature experiments are available for tungsten that tends to be brittle under these conditions, for which the most interesting finding is that the $(\bar{1}10)[\bar{1}\bar{1}\bar{1}]$ system dominates for most orientations for which $\chi>0$ although its Schmid factor is only a half of the most highly stressed $(\bar{1}01)[\bar{1}\bar{1}\bar{1}]$ system. However, this deformation mode is reminiscent of the anomalous slip observed in a number of high-purity BCC metals at very low temperatures [39-51] and it is feasible that this type of slip has the same origin as found in tungsten. Notwithstanding, in recent years, Seeger and Hollang [35-37] performed a series of detailed experiments on ultra-high purity molybdenum single crystals with the main objective to investigate the tension-compression asymmetry of the yield stress. In the following section we compare our predictions with these experimental results.

## 7. Yield stress asymmetry in tension and compression

In the experiments of Hollang et al. [35-37] on molybdenum interesting tension-compression asymmetries of the uniaxial yield stress were found at temperatures close to 123 K. For orientations of the tensile/compressive axes $[\bar{1}49]$ ($\chi = 0$) and $[\bar{5}79]$ ($\chi = +21^\circ$) the uniaxial yield stress in compression was appreciably higher than in tension. While for the $[\bar{5}79]$ axis this could be explained by the twinning-antitwinning asymmetry of shearing, these arguments fail for $\chi = 0$ when no such asymmetry exists. Moreover, for the $[\bar{1}22]$ loading axis ($\chi = +29^\circ$) the uniaxial yield stress in tension was found to be higher than in compression. Thus the tension-compression asymmetry changes character between $\chi = +21^\circ$ and $\chi = +29^\circ$ which can in no way be attributed to the twinning-antitwinning asymmetry. In the following, we employ the proposed yield criterion (5) for molybdenum to predict the tension-compression asymmetries and compare the results with the data obtained in [36]. For this purpose, we introduce the so-called strength differential

$$SD = \frac{\sigma_t - \sigma_c}{(\sigma_t + \sigma_c)/2}. \tag{6}$$

where $\sigma_t$ and $\sigma_c$ are the uniaxial yield stresses in tension and compression, respectively. For any orientation of the loading axis these yield stresses can be determined on the basis of the yield criterion (5) as described in the previous section. Performing these calculations for a variety of orientations of the tensile/compressive axes we obtain a map of the strength differential *SD* within the standard stereographic triangle. This is displayed in Fig. 5 by shading the interior of the standard triangle such that different shadings correspond to different values of *SD*.

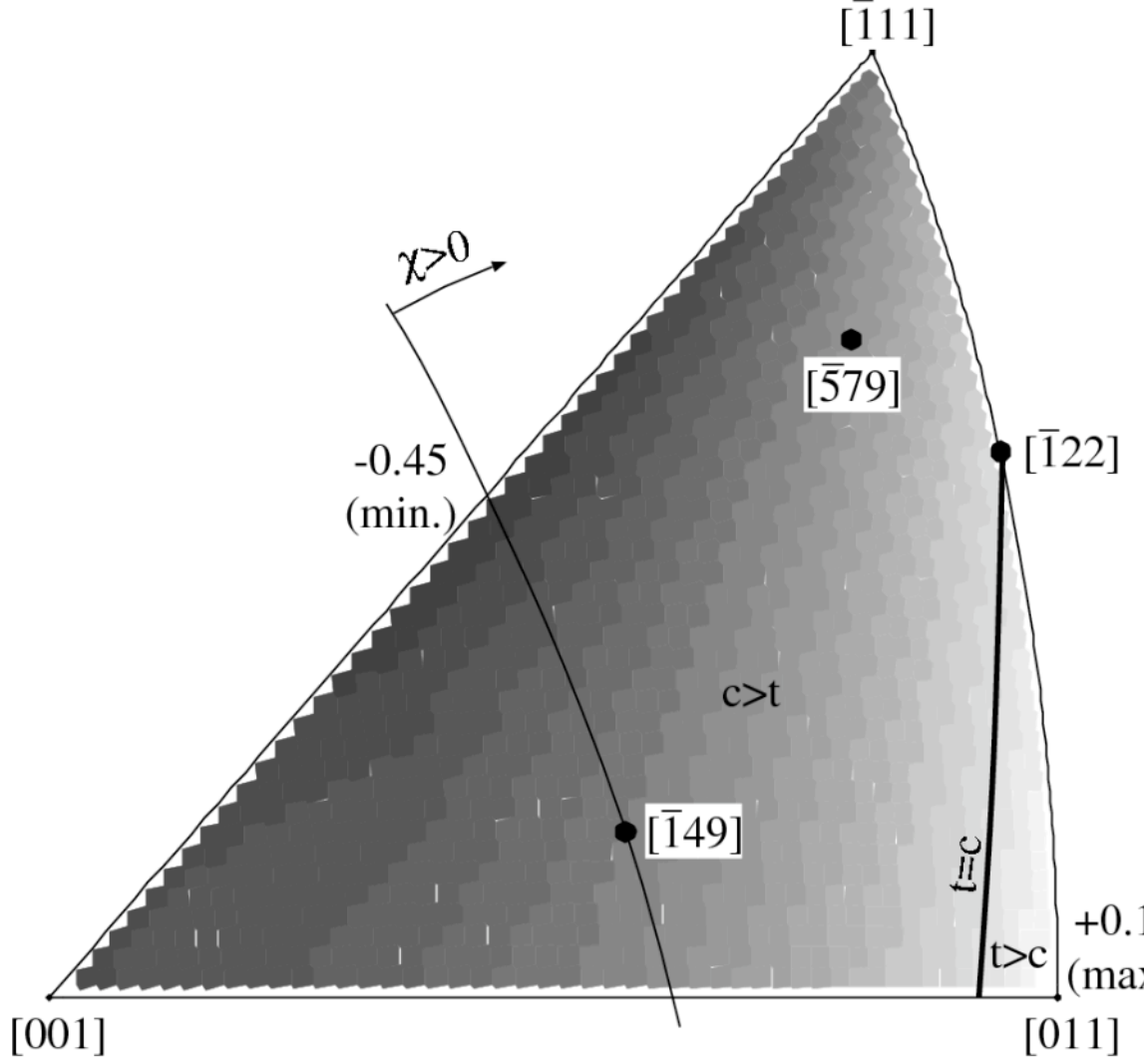

(a)

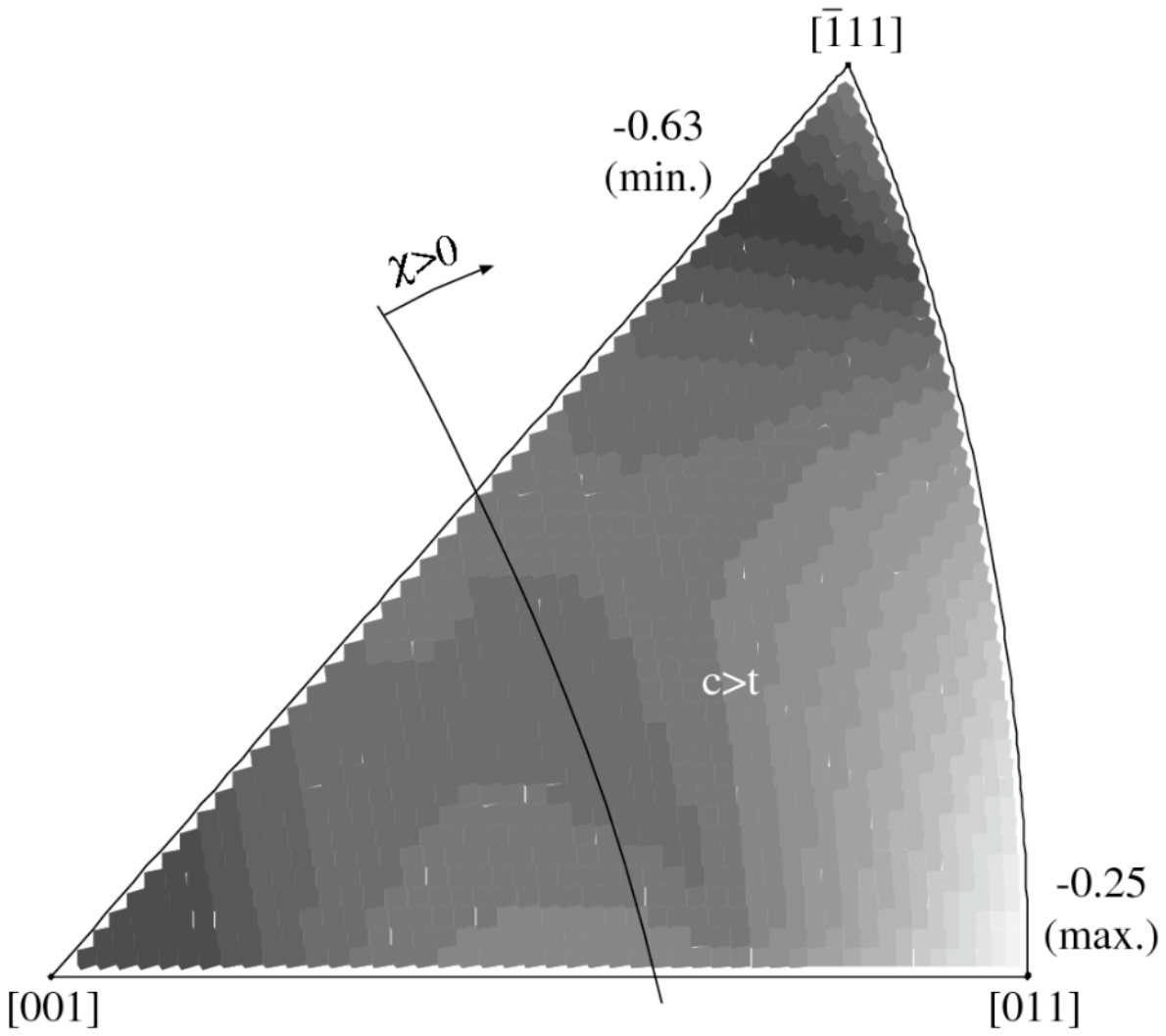

(b)

Fig. 5: Distribution of the strength differential *SD* calculated from the yield criterion (5) for molybdenum (a) and tungsten (b). Note the existence of a small region of positive *SD* in molybdenum and its lack in tungsten.

In the case of molybdenum, the yield criterion (5) predicts that for most orientations of the loading axis the uniaxial yield stress in compression is larger than in tension and thus $SD < 0$. However, for the axes approaching the $[011]$-$[\bar{1}11]$ edge, $\sigma_t$ increases relative to $\sigma_c$ and becomes equal to $\sigma_c$ for orientations along the solid curve shown in Fig. 5a. For orientations close to the [011] corner, $\sigma_t > \sigma_c$ and *SD* becomes positive. The maximum positive tension-compression asymmetry is at the [011] orientation when $SD \approx +0.11$. For comparison, the maximum negative value of the strength differential, corresponding to the loading axis in the middle of the [001]–$[\bar{1}11]$ edge, is $SD \approx -0.45$.

The above calculations of *SD* correspond to loading at 0 K since the effect of temperature has not been introduced at this stage. Nevertheless, a direct comparison between the calculated and experimentally observed strength differential [36] can be made since both $\sigma_t$ and $\sigma_c$ vary with temperature in the same way and thus *SD* is essentially temperature-independent. Such comparison is presented in Table 3 and it shows that the variation of the tension-compression asymmetry predicted by the yield criterion (5) agrees qualitatively very well with the measurements. Table 3 also contains the strength differential calculated from the restricted yield criterion ($a_2 = a_3 = 0$) that includes the effect of the twinning-antitwinning asymmetry but not the effect of shear stresses perpendicular to the slip direction. This demonstrates most emphatically that the observed tension-compression asymmetry of the yield stress is not just the consequence of the twinning-antitwinning asymmetry of shearing in the slip direction. If this were the sole reason, as has often been assumed, *SD* would be antisymmetric with respect to $\chi$ and thus would necessarily vanish at $\chi = 0$. Hence, the effect of shear stresses perpendicular to the slip direction is a major contribution to the tension-compression asymmetry.

Table 3: The strength differential *SD* for molybdenum determined experimentally in Ref. [36] and from the yield criterion (5) in the restricted ($a_2 = a_3 = 0$) and full form, respectively. In the former case only the twinning-antitwinning asymmetry of shearing can be responsible for the tension-compression asymmetry.

| | *SD* | | |
|---|---|---|---|
| | experiment | restricted criterion | full criterion |
| $[\bar{1}49]$, $\chi$=0 | –0.06 | 0 | -0.28 |
| $[\bar{5}79]$, $\chi$=+21° | –0.04 | +0.16 | -0.21 |
| $[\bar{1}22]$, $\chi$=+29° | +0.07 | +0.21 | 0.0 |

For completeness, we present in Fig. 5b the distribution of the strength differential *SD* determined from the yield criterion (5) for tungsten. In this case $SD < 0$ for all orientations of the loading axes and thus the yield stress in compression is always higher than in tension. This feature of the deformation behavior can be attributed to even stronger effect of the shear stresses perpendicular to the slip direction in tungsten than in molybdenum, which is apparent from the comparison of the coefficients $a_2$ and $a_3$ for the two metals, presented in Table 2.

## 8. Random polycrystals

The effects of non-glide stresses on the slip behavior of BCC metals have been clearly established from atomistic studies of screw dislocations and predictions of yielding in single crystals. In a related study on $L1_2$ intermetallic compounds, Qin and Bassani [29, 30] predicted a significant effect of non-glide stresses on shear localization in single crystals under both single and multiple slip. Similar effects are also predicted for BCC single crystals. We will now illustrate that the effects of non-glide stresses persist in yielding of BCC polycrystals, which also requires non-associated flow descriptions. This has been demonstrated in studies of forming limits associated with necking in thin sheets [53]. In this section, we outline a framework for an isotropic non-associated flow theory appropriate for random BCC polycrystals and in the following section cavitation instabilities are analyzed.

A classical non-associated flow relation for the plastic strain rate, $\dot{\varepsilon}^{\mathrm{p}}_{ij}$, is defined by[4]

$$\dot{\varepsilon}^{\mathrm{p}}_{ij} = \frac{\lambda}{E^{\mathrm{p}}_{\mathrm{t}}}\left(\frac{\partial F}{\partial \sigma_{kl}}\dot{\sigma}_{kl}\right)\frac{\partial G}{\partial \sigma_{ij}}, \tag{7}$$

where, $F(\sigma_{ij})$ is the yield function, $G(\sigma_{ij})$ the flow potential and $E^{\mathrm{p}}_{\mathrm{t}}$ the plastic tangent modulus. Note that the plastic strain rate is normal to the flow surface in stress space, and the term involving the normal to the yield surface ensures a smooth transition between plastic loading and elastic unloading. Here, $\lambda = 1$ for active plastic flow, i.e. when $F = F_{\mathrm{cr}}$ and $\dot{F} = \dot{F}_{\mathrm{cr}}$, and $\lambda = 0$, i.e. $\dot{\varepsilon}^{\mathrm{p}}_{ij} = 0$, if $F < F_{\mathrm{cr}}$ or if $F = F_{\mathrm{cr}}$ and $\dot{F} < \dot{F}_{\mathrm{cr}}$ where $F_{\mathrm{cr}}$ is a measure of hardness. In this study, the yield function and flow potential are taken to be homogenous of degree one in stress and, therefore, it is convenient to regard the yield function as the effective stress measure $\sigma_{\mathrm{e}}$, i.e. an active plastic state corresponds to $\sigma_{\mathrm{e}} = F_{\mathrm{cr}}$.

The plastic tangent modulus is given by the usual relation

$$E^{\mathrm{p}}_{\mathrm{t}} = \frac{\mathrm{d}\sigma_{\mathrm{e}}}{\mathrm{d}\varepsilon^{\mathrm{p}}_{\mathrm{e}}} = \frac{\mathrm{d}F_{\mathrm{cr}}}{\mathrm{d}\varepsilon^{\mathrm{p}}_{\mathrm{e}}}, \tag{8}$$

where $\varepsilon^{\mathrm{p}}_{\mathrm{e}}$ is the effective plastic strain that is defined from the usual work-equivalency argument. Since the flow potential is homogenous of degree one in stress and, therefore, for monotonic loading ($\dot{\sigma}_{\mathrm{e}} > 0$), Euler's theorem on homogenous functions implies that

$$\sigma_{ij}\dot{\varepsilon}^{\mathrm{p}}_{ij} = \frac{1}{E^{\mathrm{p}}_{\mathrm{t}}}\left(\frac{\partial F}{\partial \sigma_{kl}}\dot{\sigma}_{kl}\right)\sigma_{ij}\frac{\partial G}{\partial \sigma_{ij}} = \dot{\varepsilon}^{\mathrm{p}}_{\mathrm{e}}G(\sigma_{kl}) . \tag{9}$$

Hence, the effective plastic strain rate is defined to be conjugate to $G$, i.e.

$$\dot{\varepsilon}^{\mathrm{p}}_{\mathrm{e}} = \frac{\sigma_{ij}\dot{\varepsilon}^{\mathrm{p}}_{ij}}{G(\sigma_{kl})} . \tag{10}$$

If the plastic work rate is equated with the rate of dissipation, we note that the second law of thermodynamics requires:

$$\sigma_{ij}\dot{\varepsilon}^{\mathrm{p}}_{ij} = \frac{\lambda}{E^{\mathrm{p}}_{\mathrm{t}}}\left(\frac{\partial F}{\partial \sigma_{kl}}\dot{\sigma}_{kl}\right)\sigma_{ij}\frac{\partial G}{\partial \sigma_{ij}} = \frac{\lambda\dot{F}}{E^{\mathrm{p}}_{\mathrm{t}}}G > 0 . \tag{11}$$

Consequently, since $\lambda\dot{F} / E^{\mathrm{p}}_{\mathrm{t}}$ is always non-negative, the second law is satisfied irrespective of the functional form of $G$ if the flow function is positive and homogenous in stress.

[4] Since cavitation instabilities involve proportional stressing, we will not distinguish co-rotational time derivatives of stress from ordinary time derivatives.

In what follows, we consider isotropic behavior at the macroscopic scale and utilize recent results of a Taylor type calculation for random polycrystals, which assumes that the strain in each crystal is the same as the macroscopic strain (see e.g. [32, 33]). The yield and flow surfaces predicted from the Taylor type calculation, based upon the yield criterion (5) with the $a_i$ derived from atomistic simulations for molybdenum, are plotted as circles in Fig. 6 for plane states of stress. A family of similar surfaces is obtained as the magnitudes of the non-glide stress coefficients, $a_i$ in (2), are varied. The flow surfaces, on the other hand, have little dependence on the non-glide stresses and, therefore, are indistinguishable from the yield surface for associated flow behavior [54]. Finally, we note that we have considered both rate-dependent and rate-independent flow descriptions and, in the latter case, both with smooth yield and flow surfaces and with the effects of corners on the flow surface [53]. Here we limit our attention to smooth yield surfaces and adopt a flow rule of the form of (7).

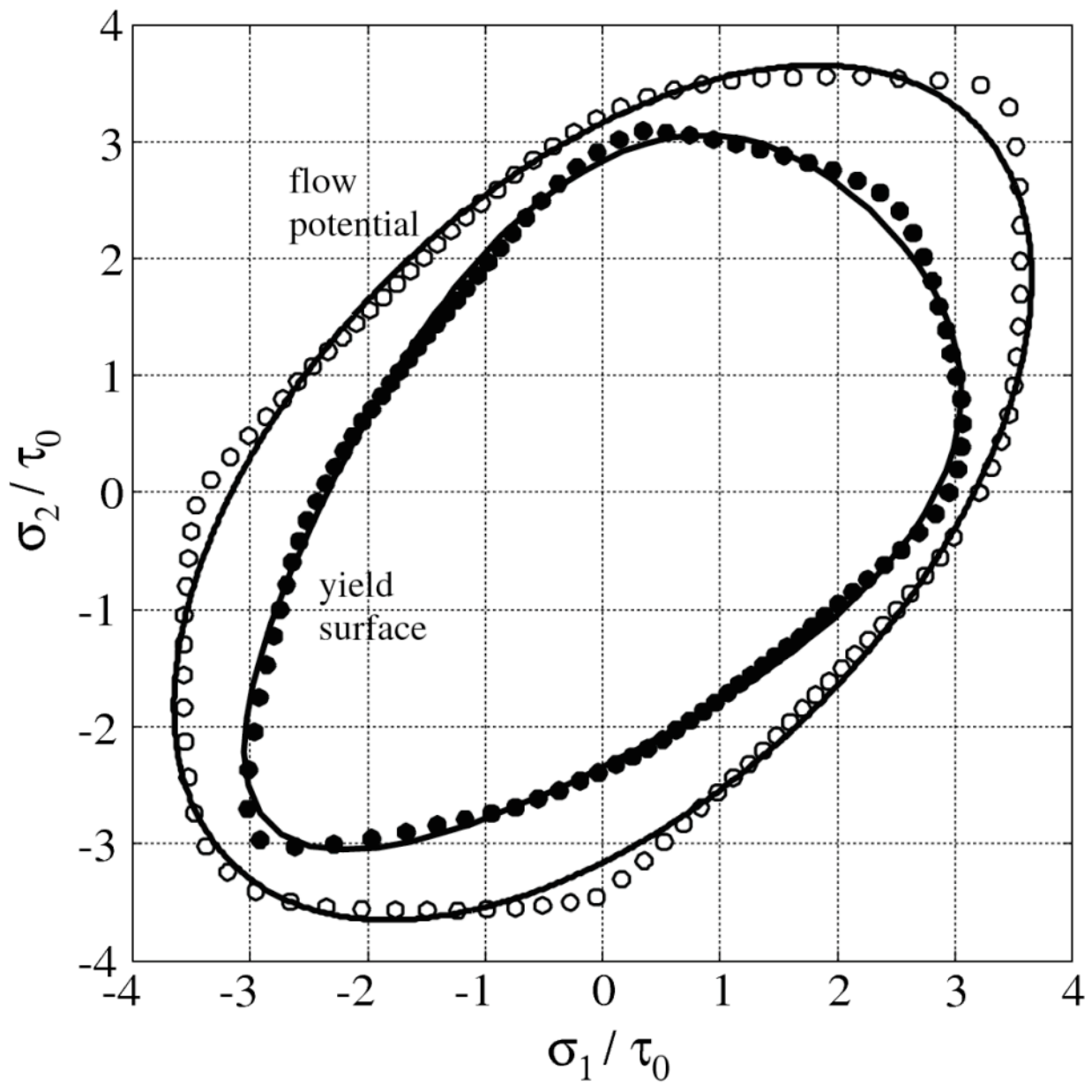

Fig. 6: Yield and flow surfaces predicted from a Taylor model for BCC Mo (circles) and best fits: $F$ with $b$=–0.72 in (12a) and $G$ (12b).

Simple isotropic functions that accurately describe computed yield and flow surfaces for random BCC polycrystals obtained from Taylor calculations, such as those shown in Fig. 6, are given by:

$$F = \sqrt{3}\left[\left(J_2\right)^{3/2} + bJ_3\right]^{1/3} \tag{12a}$$

$$G = \sqrt{3J_2}\ , \tag{12b}$$

where $J_2 = s_{kl}s_{kl}/2$ and $J_3 = s_{ij}s_{jk}s_{ki}/3$ are the second and third invariants of the deviatoric Cauchy stress $s_{ij}$, respectively. In this simple isotropic theory, the parameter $b$ entering the yield function (12a) is the non-associated flow parameter and the only measure of the effects of non-glide stresses. Note that the yield function $F$ reduces to the classical von Mises function (12b) for $b$=0. These functions are plotted as the continuous curves in Fig. 6, with $b = -0.72$ the least-square best fit to the yield surface computed for single crystals governed by the yield criterion (5) with the coefficients of non-glide stresses chosen from the atomistic results for Mo. With $J_2$ and $J_3$ computed for uniaxial stress states, the strength differential (6) is given in terms of $b$ and using (12a) it can be written as

$$SD = 2\frac{\left(1-2b/3\sqrt{3}\right)^{1/3} - \left(1+2b/3\sqrt{3}\right)^{1/3}}{\left(1-2b/3\sqrt{3}\right)^{1/3} + \left(1+2b/3\sqrt{3}\right)^{1/3}} \approx -0.257\, b \ . \tag{13}$$

It is worth noting that the best fit to the flow surface in Fig. 6 with a function of the form of (12a) gives $b \approx 0$, which is the von Mises function (12b).

In closing this section, we note that non-associated flow models also have been adopted to describe frictional materials, e.g. granular and geological materials [55-57], and to account for pressure-sensitive yield in metals [58-61]. With these material behaviors in mind, the effects of non-associated flow on strain localization have been investigated, while predictions have been mostly restricted to shear bands in plane strain tension [55, 62]. Kuroda [60] has studied sheet necking and has shown that the effect of non-associated flow arising from pressure sensitivity on forming limits is small. In contrast, we have found [53] a significant effect of non-associated flow on sheet forming limits from calculations that utilize the isotropic yield and flow functions (12a) and (12b), respectively, that approximate well the Taylor surfaces for random BCC polycrystals.

## 9. Cavitation instabilities

The fact that non-glide stress effects persist at the level of polycrystals and can significantly alter the failure criteria is readily demonstrated in the problem of cavitation instabilities. These instabilities arise when the energy available from the strained material surrounding the cavity is sufficient to drive continued expansion (see e.g. [63]). The problem addressed here is defined as follows: a spherically symmetric stress (i.e. hydrostatic tension) is applied at infinity and conditions are sought for the critical value of this stress when a spherical cavity grows without bounds, i.e. when the radius of the cavity approaches infinity in the deformed configuration. The existence of cavitation instabilities in elastic-plastic solids was recognized by Bishop et al. [64] and Hill [52] also determined limit states for the equivalent problems of cavities subjected to internal pressure. Huang et al. [65] studied cavitation instabilities under axisymmetric states of remote stress in elastic-plastic solids.

In this analysis, we assume a rigid-plastic rate-independent material behavior for simplicity (i.e. the elastic strains are neglected). Furthermore, we assume that the loading is monotonic and the material strain hardens, in which case the plastic tangent modulus is always positive. Let $R$ and $r$ denote the radial coordinates of material points, measured from the center of a spherical cavity, in the undeformed and deformed configurations, respectively. The principal stretches along the radial, zenithal, and azimuthal directions are given, respectively, by

$$\lambda_r = \left(\frac{R}{r}\right)^2, \quad \lambda_\theta = \lambda_\phi = \frac{r}{R} \tag{14}$$

and the corresponding radial strain rate is

$$\dot{\varepsilon}_{rr} = \frac{\dot{\lambda}_r}{\lambda_r} = -2\frac{\dot{r}}{r} \ . \tag{15}$$

Since hydrostatic tension is applied at infinity, $\dot{r} > 0$ and, therefore, $\dot{\varepsilon}_{rr} < 0$. For monotonic loading, stressing is everywhere proportional and, therefore, from (7) with (12a) and (12b) the radial strain rate is

$$\dot{\varepsilon}_{rr} = \frac{3}{2E_t^p}\dot{F}\frac{s_{rr}}{G} \ . \tag{16}$$

From the above expression and the fact that the loading is monotonic it is straightforward to show that $\dot{F} > 0$, and $s_{rr} < 0$. Thus, the effective stress reduces to

$$\sigma_e = \frac{3}{2}\left|s_{rr}\right|\left(\frac{3\sqrt{3} - 2b}{3\sqrt{3} + 2b}\right)^{1/3} = F_{\mathrm{cr}}(\varepsilon_e^p) \ . \tag{17}$$

Since elastic strains are neglected in this analysis, it can be shown from the flow relation (7) that the effective plastic strain is

$$\varepsilon_{\mathrm{e}}^{\mathrm{p}} = \int \sqrt{\frac{2}{3}\dot{\varepsilon}_{ij}\dot{\varepsilon}_{ij}}\,\mathrm{d}t = \int \sqrt{\frac{2}{3}\left(\dot{\varepsilon}_{rr}^2 + \dot{\varepsilon}_{\theta\theta}^2 + \dot{\varepsilon}_{\phi\phi}^2\right)}\,\mathrm{d}t = 2\ \ln\left(\frac{r}{R}\right) \ . \tag{18}$$

Therefore, the equilibrium condition for spherically symmetric deformations leads to

$$\frac{\mathrm{d}\sigma_{rr}}{\mathrm{d}r} = \frac{-3s_{rr}}{r} = 2\left(\frac{3\sqrt{3} + 2b}{3\sqrt{3} - 2b}\right)^{1/3}\frac{F_{\mathrm{cr}}\left(2\ln(r/R)\right)}{r} \ . \tag{19}$$

For the case of hydrostatic tension applied at infinity, i.e.

$$\sigma_{rr} = 0 \ \text{ at } \ r = r_{\mathrm{c}} \ \text{ and } \ \sigma_{rr} = \sigma^\infty \ \text{ as } \ r \to \infty \ , \tag{20}$$

the relation between the current radius of the cavity, $r_{\mathrm{c}}$, and the stress at infinity is given by

$$\sigma^\infty = 2\left(\frac{3\sqrt{3} + 2b}{3\sqrt{3} - 2b}\right)^{1/3}\int_{r_{\mathrm{c}}}^{\infty}\frac{F_{\mathrm{cr}}\left(2\ln(r/R)\right)}{r}\,\mathrm{d}r \ . \tag{21}$$

With the incompressibility constraint for the material surrounding the cavity, we can write

$$\frac{r}{R} = \left(1 - \frac{r_{\mathrm{c}}^3 - R_{\mathrm{c}}^3}{r^3}\right)^{1/3} , \tag{22}$$

where $R_{\mathrm{c}}$ is the initial radius of the cavity. The cavitation limit stress $\sigma_{\mathrm{crit}}$ follows from (21) by setting $R_{\mathrm{c}}/r_{\mathrm{c}} \to 0$, i.e.

$$\sigma_{\text{crit}} = 2\left(\frac{3\sqrt{3}+2b}{3\sqrt{3}-2b}\right)^{1/3}\int_1^{\infty} F_{\text{cr}}\left[-\frac{2}{3}\ln\left(1-\eta^{-3}\right)\right]\eta^{-1}\text{d}\eta \ , \tag{23}$$

where $\eta = r/r_c$ .

Here a class of power-law hardening materials is considered with

$$F_{\text{cr}}\left(\varepsilon_{\text{e}}^{\text{p}}\right) = \begin{cases} \sigma_{\text{o}}\,\varepsilon_{\text{e}}^{\text{p}}/\varepsilon_{\text{o}} & \text{if } \varepsilon_{\text{e}}^{\text{p}} \le \varepsilon_{\text{o}} \\ \sigma_{\text{o}}\left(\dfrac{\varepsilon_{\text{e}}^{\text{p}}}{\varepsilon_{\text{o}}}\right)^{N} & \text{if } \varepsilon_{\text{e}}^{\text{p}} \ge \varepsilon_{\text{o}} \end{cases} , \tag{24}$$

where $\sigma_{\text{o}}$ , $\varepsilon_{\text{o}}$ and $N$ are material constants. For this power-law hardening behavior, the estimate for the cavitation limit stress (23) becomes

$$\sigma_{\text{crit}} = 2\left(\frac{3\sqrt{3}+2b}{3\sqrt{3}-2b}\right)^{1/3}\left\{\begin{array}{c} \displaystyle\int_1^{\eta_{\text{o}}} \frac{\sigma_{\text{o}}}{\left(\varepsilon_{\text{o}}\right)^{N}}\left[-\frac{2}{3}\ln\left(1-\eta^{-3}\right)\right]^{N}\eta^{-1}\text{d}\eta \ + \\ \displaystyle\int_{\eta_{\text{o}}}^{\infty} -\frac{2}{3}\frac{\sigma_{\text{o}}}{\varepsilon_{\text{o}}}\ln\left(1-\eta^{-3}\right)\eta^{-1}\text{d}\eta \end{array}\right\} , \tag{25}$$

where

$$\eta_{\text{o}} = \left(1-e^{-3\varepsilon_{\text{o}}/2}\right)^{-1/3} . \tag{26}$$

The cavitation limit versus the hardening exponents $N$ is plotted in Fig. 7 for various values of the strength differential, *SD*, which is given in terms of the coefficient $b$ in (13). As in the case of associated flow [65], the limit stress increases with increasing hardening exponent even for non-associated flow. In addition, there is a significant dependence on *SD*, as seen in Fig. 8 for $N = 0.1$. Recall that for a random polycrystal with the single crystal yield criterion derived from the atomistic results for molybdenum $SD \approx 0.2$ . For that value of the strength differential there is almost 20% reduction in the cavitation limit.

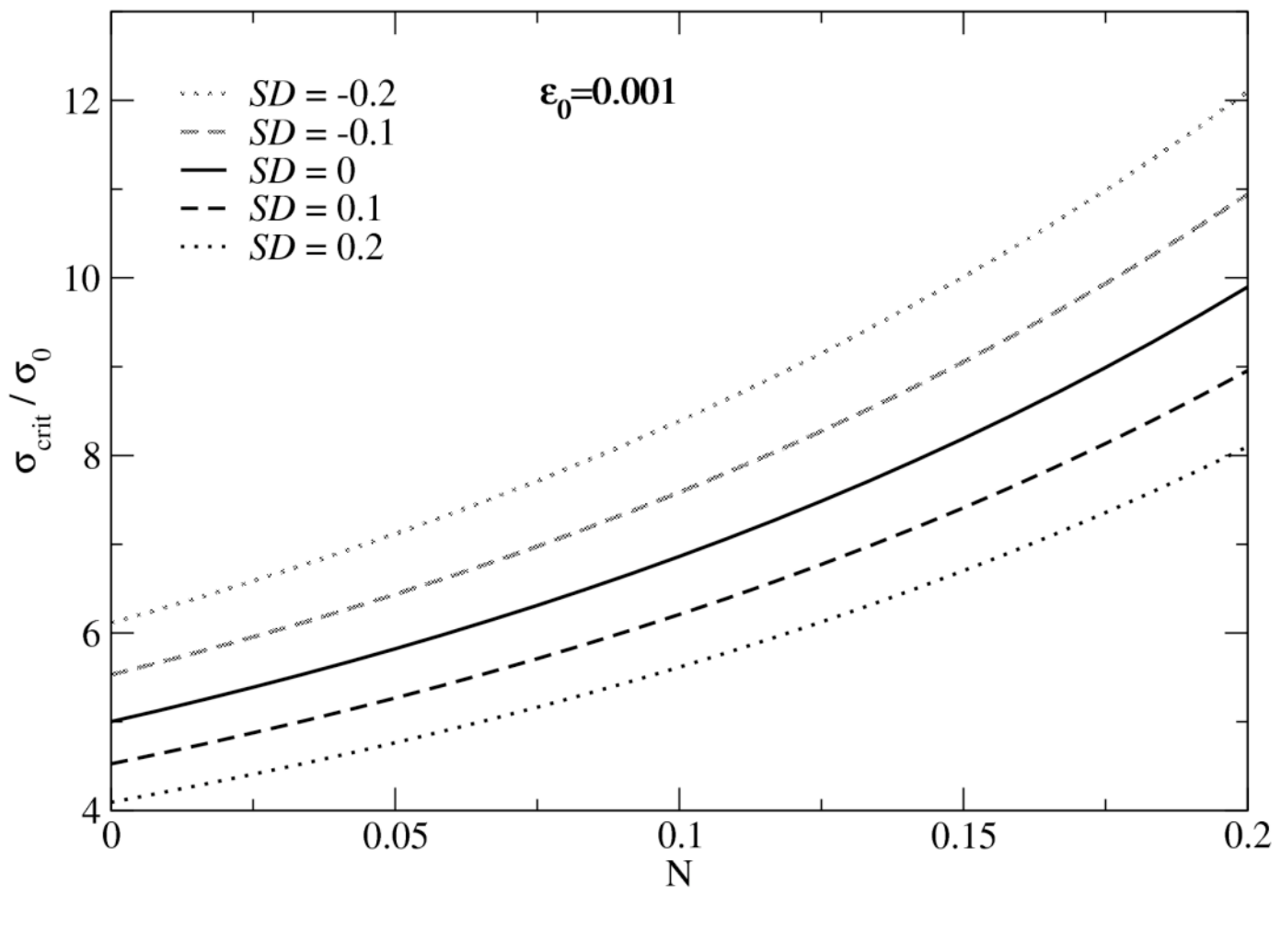

Fig. 7: Cavitation limits versus hardening exponent $N$ for various values of strength differential *SD*.

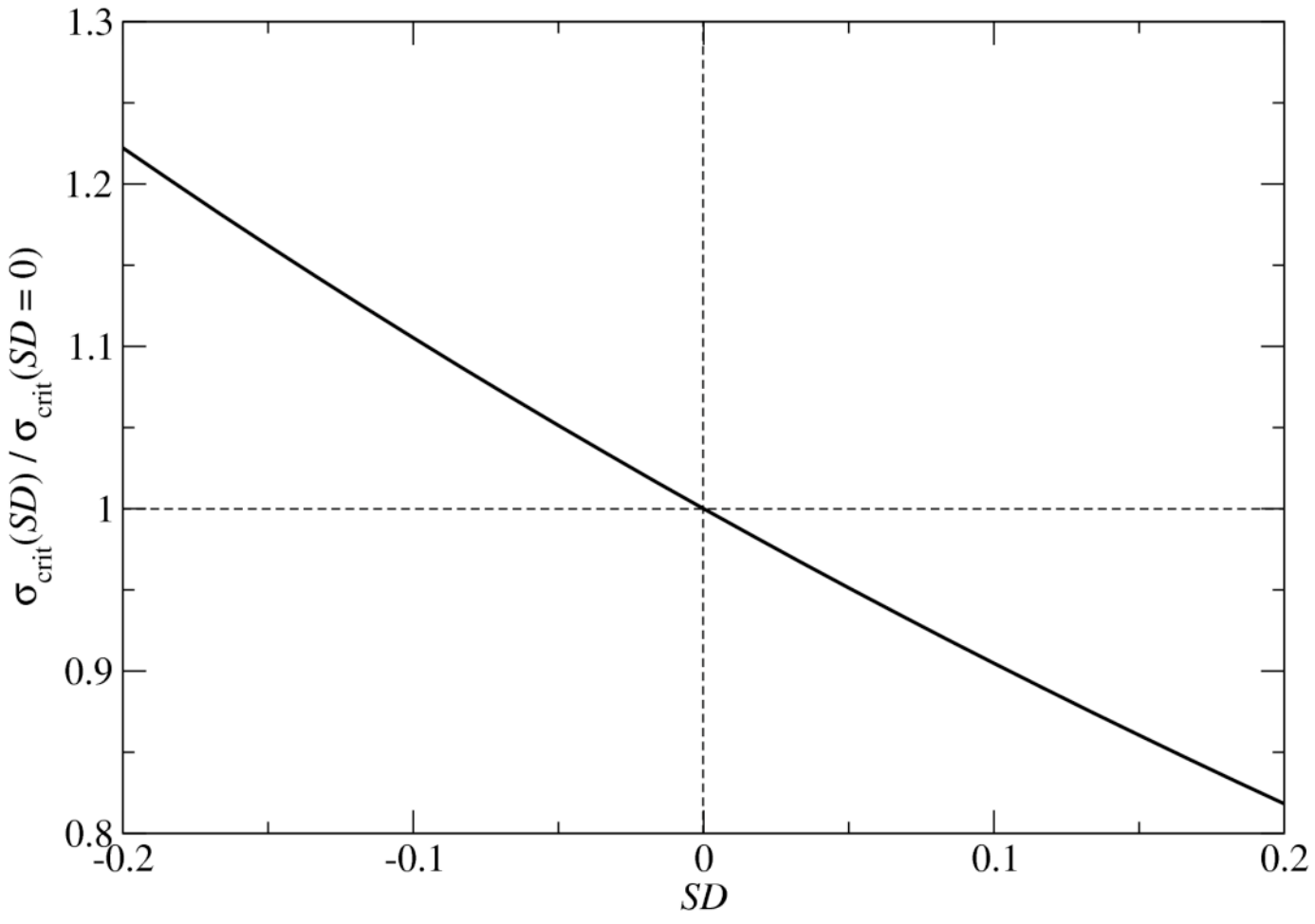

Fig. 8: Variation of the cavitation stress normalized by the result for associated flow ( $SD = 0$ ) with the strength differential $SD$ for the hardening coefficient $N = 0.1$.

## 10. Conclusions

Based on atomistic studies of the glide of an isolated 1/2[111] screw dislocation, presented in the Part I of this series of papers, we constructed the analytical yield criteria for Mo and W that reflect the effect of non-glide stresses on the CRSS (Peierls stress) for glide of these dislocations. In these criteria, formulated following the original development in [29, 30], a linear combination of two shear stresses parallel to and two shear stresses perpendicular to the slip direction has to reach a critical value. Only the shear stress parallel to the slip direction acting in the {110} type glide plane, called the Schmid stress, does work as the dislocation glides. The other three shear stresses affect the magnitude of the CRSS by modifying the structure of the dislocation core and their effect causes the breakdown of the Schmid law. In general, when plastic deformation is affected by components of the applied stress tensor other than the Schmid stress, such behavior is defined as a non-associated plastic flow. In the setting of continuum mechanics it means that the yield function and the flow potential are not equal. The physical origin of this non-associated flow behavior is the complex response of the non-planar cores of 1/2⟨111⟩ screw dislocations to a general state of stress.

The yield criterion formulated in this paper contains only four adjustable parameters that are readily determined using the results of the atomistic calculations presented in Part I. Specifically, the dependence of the CRSS (Peierls stress) on the orientation of the MRSSP, i.e. the angle $\chi$, is used to parameterize the dependence of the CRSS on the shear stress parallel to the slip direction in a {110} plane other than the glide plane. This reproduces the twinning-antitwinning asymmetry of the dislocation glide and determines parameters $a_1$ and $\tau^*_{cr}$ in (2) and (5). Subsequently, the atomistically calculated dependence of the CRSS on the shear stress perpendicular to the slip direction, $\tau$, is accurately reproduced via parameters $a_2$ and $a_3$ in (2) and (5).

Since all {110}⟨111⟩ slip systems are equivalent, the tensorial representation of the criterion (5) can be directly employed in crystal plasticity studies. For example, it can be utilized to determine the first system to operate under a given applied loading and to assess the slip activity of each system during continued plastic deformation. We have shown for tensile loading along various axes within the standard stereographic triangle that both Mo and W deform predominantly by slip on the most highly stressed $(\bar{1}01)[111]$ system. On the other hand, in compression the slip activity depends both on the material and on the orientation of the loading axis. For a broad range of orientations the most operative slip system in Mo is still the most highly stressed $(\bar{1}01)[\bar{1}\,\bar{1}\,\bar{1}]$ system but in W it is the $(\bar{1}10)[\bar{1}\,\bar{1}\,\bar{1}]$ system. To the best of our knowledge, there are no experimental data to verify or refute the latter prediction. For orientations of loading axes close to the corners and sides of the standard stereographic triangle the calculations predict a simultaneous operation of two or more slip systems in both Mo and W, i.e. the multislip that has been frequently observed. Moreover, in W slip is predicted to occur on a low-stressed $(\bar{1}10)[\bar{1}\,\bar{1}\,\bar{1}]$ system for the loading axes corresponding to $\chi > 0$ (Fig. 4b), while in Mo the same slip mode is observed only for orientations close to $\chi = 30^\circ$ (Fig. 4a). In both metals, this is reminiscent of the anomalous slip observed in many high-purity BCC metals at low temperature.

A direct comparison between experimental observations and predictions based on the constructed yield criterion can be made for the tension-compression asymmetry of the yield stress that was measured in molybdenum in [35-37]. For Mo the yield criterion predicts that the critical compressive stress is larger than the critical tensile stress for a broad range of orientations of the loading axis within the standard stereographic triangle. The character of this asymmetry reverses only for orientations close to the [011] corner where the yield stress in tension becomes higher than that in compression. This finding is in an excellent qualitative agreement with the observations in [35-37]. An interesting feature of this asymmetry is that it is significant for the $[\bar{1}49]$ loading axis when the $(\bar{1}01)$ plane is the MRSSP and $\chi = 0$. If only the shear stress parallel to the slip direction played role the asymmetry would necessarily vanish for this orientation since such stress can induce an asymmetry, known as the twinning-antitwinning asymmetry, only for orientations of loading with $\chi \neq 0$. This disparity was noted in Ref. [36] and attributed to "a modification of the Peierls potential by stress components other than the resolved shear stress". In the setting of the non-associated plastic flow model, we identify this contribution as an effect of shear stresses perpendicular to the slip direction. For other orientations of the loading axis the asymmetry is the result of combination of twinning-antitwinning asymmetry and effects of stresses perpendicular to the slip direction.

For W the yield criterion predicts that the critical compressive stress is larger than the critical tensile stress for any orientation of the loading axis within the standard stereographic triangle. Since the twinning-antitwinning asymmetry is negligible in W the predicted tension-compression asymmetry is entirely due to the shear stresses perpendicular to the slip direction. Unfortunately, no experimental observations that would test this finding are currently available.

This paper concludes with predictions of the effects of non-glide stresses at the polycrystalline level, in which individual grains undergo multiple slip. Using the slip system yield criteria derived from the atomistic studies and a Taylor homogenization procedure for a random polycrystal we have demonstrated that the effects of non-planar cores of screw dislocations persist at the polycrystal level. The polycrystalline constitutive behavior also corresponds to non-associated flow in which the yield and flow functions depend only on deviatoric components of stress. Several earlier continuum analyses have found that the effects of non-glide stresses significantly affect critical phenomena at macroscopic scales such as shear localization in single crystals [29, 30]. In this paper, we demonstrated this fact on the problem of cavitation in a ductile plastic solid. The dependence of the cavitation limit on the strength differential (*SD*), which is a convenient measure of the extent of non-associated flow, is shown to scale roughly in proportion to *SD* (see Fig. 7). For the strength differential derived from the yield criterion for molybdenum there is almost 20% reduction in the cavitation limit. In another related study, Racherla and Bassani [52] have shown that non-associated flow plays an important role also in forming limits of biaxially stretched sheets. Overall, these findings demonstrate that the effects of non-planar core structure of screw dislocations in BCC metals and related complex influence of applied loading on dislocation glide persist in macroscopic deformation of both single and polycrystals.

**Acknowledgements**

This research was supported by the Department of Energy, BES Grant no. DE-PG02-98ER45702 (RG and VV), the NSF Grant no. DMR 02-19243, and by DOE/ASCI through Lawrence Livermore National Laboratory (VR and JLB).